\documentclass[twocolumn,rmp]{revtex4}

\usepackage{dcolumn}
\usepackage{amsmath}

\usepackage{graphicx}

\renewcommand{\d}{\mathrm{d}}

\newcommand{\half}{\mbox{$\frac12$}}
\newcommand{\set}[1]{\lbrace#1\rbrace}
\newcommand{\av}[1]{\langle#1\rangle}
\newcommand{\etal}{{\it{}et~al.}}
\newcommand{\defn}{\textit}
\renewcommand{\vec}{\mathbf}
\newcommand{\mat}{\mathbf}
\newcommand{\Tr}{\mathrm{Tr}}

\newcommand{\Ord}{\mathrm{O}}

\bibpunct{[}{]}{,}{n}{}{,}

\begin{document}

\title{Finding community structure in networks using the eigenvectors
of matrices}
\author{M. E. J. Newman}
\affiliation{Department of Physics and Center for the Study of Complex
Systems, University of Michigan, Ann Arbor, MI 48109--1040}
\begin{abstract}
We consider the problem of detecting communities or modules in networks,
groups of vertices with a higher-than-average density of edges connecting
them.  Previous work indicates that a robust approach to this problem is
the maximization of the benefit function known as ``modularity'' over
possible divisions of a network.  Here we show that this maximization
process can be written in terms of the eigenspectrum of a matrix we call
the modularity matrix, which plays a role in community detection similar to
that played by the graph Laplacian in graph partitioning calculations.
This result leads us to a number of possible algorithms for detecting
community structure, as well as several other results, including a spectral
measure of bipartite structure in networks and a new centrality measure
that identifies those vertices that occupy central positions within the
communities to which they belong.  The algorithms and measures proposed are
illustrated with applications to a variety of real-world complex networks.
\end{abstract}
\maketitle

\section{Introduction}
Networks have attracted considerable recent attention in physics and other
fields as a foundation for the mathematical representation of a variety of
complex systems, including biological and social systems, the Internet, the
worldwide web, and many others~\cite{Newman03d,Boccaletti06,DM03b,NBW06}.
A common feature of many networks is ``community structure,'' the tendency
for vertices to divide into groups, with dense connections within groups
and only sparser connections between them~\cite{GN02,Newman04b}.  Social
networks~\cite{GN02}, biochemical networks~\cite{HHJ03,GA05,PDFV05}, and
information networks such as the web~\cite{FLGC02}, have all been shown to
possess strong community structure, a finding that has substantial
practical implications for our understanding of the systems these networks
represent.  Communities are of interest because they often correspond to
functional units such as cycles or pathways in metabolic
networks~\cite{GA05,PDFV05,HH06} or collections of pages on a single topic
on the web~\cite{FLGC02}, but their influence reaches further than this.  A
number of recent results suggest that networks can have properties at the
community level that are quite different from their properties at the level
of the entire network, so that analyses that focus on whole networks and
ignore community structure may miss many interesting features.

For instance, in some social networks one finds individuals with different
mean numbers of contacts in different groups; the individuals in one group
might be gregarious, having many contacts with others, while the
individuals in another group might be more reticent.  An example of this
behavior is seen in networks of sexual contacts, where separate communities
of high- and low-activity individuals have been
observed~\cite{Garnett96,Aral99}.  If one were to characterize such a
network by quoting only a single figure for the average number of contacts
an individual has, one would be missing features of the network directly
relevant to questions of scientific interest such as epidemiological
dynamics~\cite{GAM89}.

It has also been shown that vertices' positions within communities can
affect the role or function they assume.  In social networks, for example,
it has long been accepted that individuals who lie on the boundaries of
communities, bridging gaps between otherwise unconnected people, enjoy an
unusual level of influence as the gatekeepers of information flow between
groups~\cite{Granovetter73,Burt76,Freeman77}.  A surprisingly similar
result is found in metabolic networks, where metabolites that straddle the
boundaries between modules show particular persistence across
species~\cite{GA05}.  This finding might indicate that modules in metabolic
networks possess some degree of functional independence within the cell,
allowing vertices central to a module to change or disappear with
relatively little effect on the rest of the network, while vertices on the
borders of modules are less able to change without affecting other aspects
of the cellular machinery.

One can also consider the communities in a network themselves to form a
higher level meta-network, a coarse-grained representation of the full
network.  Such coarse-grained representations have been used in the past as
tools for visualization and analysis~\cite{NG04} but more recently have
also been investigated as topologically interesting entities in their own
right.  In particular, networks of modules appear to have degree
distributions with interesting similarities to but also some differences
from the degree distributions of other networks~\cite{PDFV05}, and may also
display so-called preferential attachment in their formation~\cite{PPV06},
indicating the possibility of distinct dynamical processes taking place at
the level of the modules.

For all of these reasons and others besides there has been a concerted
effort in recent years within the physics community and elsewhere to
develop mathematical tools and computer algorithms to detect and quantify
community structure in networks.  A huge variety of community detection
techniques have been developed, based variously on centrality measures,
flow models, random walks, resistor networks, optimization, and many other
approaches~\cite{GN02,Zhou03,Krause03,WH04a,Radicchi04,NG04,Newman04a,CSCC04,FLM04,RB04,DM04,ZL04,PDFV05,GA05,Clauset05,PL05,Newman06b,DDA06,RB06,Hastings06}.
For reviews see Refs.~\cite{Newman04b,DDDA05}.

In this paper we focus on one approach to community detection that has
proven particularly effective, the optimization of the benefit function
known as ``modularity'' over the possible divisions of a network.  Methods
based on this approach have been found to produce excellent results in
standardized tests~\cite{DDDA05,GLH06}.  Unfortunately, exhaustive
optimization of the modularity demands an impractically large computational
effort, but good results have been obtained with various approximate
optimization techniques, including greedy
algorithms~\cite{Newman04a,CNM04}, simulated annealing~\cite{GSA04,RB06},
and extremal optimization~\cite{DA05}.  In this paper we describe a
different approach, in which we rewrite the modularity function in matrix
terms, which allows us to express the optimization task as a spectral
problem in linear algebra.  This approach leads to a family of fast new
computer algorithms for community detection that produce results
competitive with the best previous methods.  Perhaps more importantly, our
work also leads to a number of useful insights about network structure via
the close relations we will demonstrate between communities and matrix
spectra.

Our work is by no means the first to find connections between divisions of
networks and matrix spectra.  There is a large literature within computer
science on so-called spectral partitioning, in which network properties are
linked to the spectrum of the graph Laplacian
matrix~\cite{Fiedler73,PSL90,Fjallstrom98}.  This method is different from
the one introduced here and is not in general well suited to the problem of
community structure detection.  The reasons for this, however, turn out to
be interesting and instructive, so we begin our presentation with a brief
review of the traditional spectral partitioning method in
Section~\ref{specpart}.  A consideration of the deficiencies of this method
in Section~\ref{modsec} leads us in Sections~\ref{specmod}--\ref{multiway}
to introduce and develop at length our own method, which is based on the
characteristic matrix we call the ``modularity matrix.''
Sections~\ref{negative} and~\ref{otheruses} explore some further ideas
arising from the study of the modularity matrix but not directly related to
community detection.  In Section~\ref{concs} we give our conclusions.  A
brief report of some of the results described in this paper has appeared
previously as Ref.~\cite{Newman06b}.

\section{Graph partitioning and the Laplacian matrix}
\label{specpart}
There is a long tradition of research in computer science on graph
partitioning, a problem that arises in a variety of contexts, but most
prominently in the development of computer algorithms for parallel or
distributed computation.  Suppose a computation requires the performance of
some number~$n$ of tasks, each to be carried out by a separate process,
program, or thread running on one of $c$ different computer processors.
Typically there is a desired number of tasks or volume of work to be
assigned to each of the processors.  If the processors are identical, for
instance, and the tasks are of similar complexity, we may wish to assign
the same number of tasks to each processor so as to share the workload
roughly equally.  It is also typically the case that the individual tasks
require for their completion results generated during the performance of
other tasks, so tasks must communicate with one another to complete the
overall computation.  The pattern of required communications can be thought
of as a network with $n$ vertices representing the tasks and an edge
joining any pair of tasks that need to communicate, for a total of $m$
edges.  (In theory the amount of communication between different pairs of
tasks could vary, leading to a \emph{weighted} network, but we here
restrict our attention to the simplest unweighted case, which already
presents interesting challenges.)

Normally, communications between processors in parallel computers are slow
compared to data movement within processors, and hence we would like to
keep such communications to a minimum.  In network terms this means we
would like to divide the vertices of our network (the processes) into
groups (the processors) such that the number of edges between groups is
minimized.  This is the graph partitioning problem.

Problems of this type can be solved exactly in polynomial time~\cite{GH88},
but unfortunately the polynomial in question is of leading order~$n^{c^2}$,
which is already prohibitive for all but the smallest networks even when
$c$ takes the smallest possible value of~2.  For practical applications,
therefore, a number of approximate solution methods have been developed
that appear to give reasonably good results.  One of the most widely used
is the spectral partitioning method, due originally to
Fiedler~\cite{Fiedler73} and popularized particularly by
Pothen~\etal~\cite{PSL90}.  We here consider the simplest instance of the
method, where $c=2$, i.e.,~where our network is to be divided into just two
non-intersecting subsets such that the number of edges running between the
subsets is minimized.

We begin by defining the adjacency matrix~$\mat{A}$ to be the matrix with
elements
\begin{equation}
A_{ij} = \begin{cases}
           \enspace 1 & \text{if there is an edge joining vertices $i,j$,} \\
           \enspace 0 & \text{otherwise.}
         \end{cases}
\label{adjacency}
\end{equation}
(We restrict our attention in this paper to undirected networks, so that
$\mat{A}$ is symmetric.)  Then the number of edges~$R$ running between our
two groups of vertices, also called the \defn{cut size}, is given by
\begin{equation}
R = \half\!\!
    \sum_{\parbox{3em}{\scriptsize\centering $i,j$ in different groups}}\!\!
    A_{ij},
\label{cutsize1}
\end{equation}
where the factor of $\half$ compensates for our counting each edge twice in
the sum.

To put this in a more convenient form, we define an \defn{index
vector}~$\vec{s}$ with $n$ elements
\begin{equation}
s_i = \begin{cases}
        \enspace +1 & \text{if vertex~$i$ belongs to group~1,} \\
        \enspace -1 & \text{if vertex~$i$ belongs to group~2.}
      \end{cases}
\label{defss}
\end{equation}
(Note that $\vec{s}$ satisfies the normalization condition
$\vec{s}^T\vec{s}=n$.)  Then
\begin{equation}
\half(1-s_is_j) = \begin{cases}
                    \enspace 1 & \text{if $i$ and $j$ are in different groups,} \\
                    \enspace 0 & \text{if $i$ and $j$ are in the same group,}
                  \end{cases}
\label{sisj}
\end{equation}
which allows us to rewrite Eq.~\eqref{cutsize1} as
\begin{equation}
R = \mbox{$\frac14$} \sum_{ij} (1-s_is_j) A_{ij}.
\label{cutsize2}
\end{equation}
Noting that the number of edges~$k_i$ connected to a vertex~$i$---also
called the
\defn{degree} of the vertex---is given by
\begin{equation}
k_i = \sum_j A_{ij},
\label{degree}
\end{equation}
the first term of the sum in~\eqref{cutsize2} is
\begin{equation}
\sum_{ij} A_{ij} = \sum_i k_i = \sum_i s_i^2 k_i
  = \sum_{ij} s_i s_j k_i \delta_{ij},
\end{equation}
where we have made use of $s_i^2=1$ (since $s_i=\pm1$), and $\delta_{ij}$
is 1 if $i=j$ and zero otherwise.  Thus
\begin{equation}
R = \mbox{$\frac14$} \sum_{ij} s_i s_j (k_i \delta_{ij} - A_{ij}).
\end{equation}
We can write this in matrix form as
\begin{equation}
R = \mbox{$\frac14$} \vec{s}^T\mat{L}\vec{s},
\label{defsr}
\end{equation}
where $\mat{L}$ is the real symmetric matrix with elements $L_{ij}=k_i
\delta_{ij} - A_{ij}$, or equivalently\footnote{We assume here that the
network is a \defn{simple graph}, having at most one edge between any pair
of vertices and no self-edges (edges that connect vertices to themselves).}
\begin{equation}
L_{ij} = \begin{cases}
           k_i & \text{if $i=j$,} \\
           -1  & \text{if $i\ne j$ and there is an edge $(i,j)$,} \\
           0   & \text{otherwise.}
         \end{cases}
\end{equation}
$\mat{L}$ is called the \defn{Laplacian matrix} of the graph or sometimes
the \defn{admittance matrix}.  It appears in many contexts in the theory of
networks, such as the analysis of diffusion and random walks on
networks~\cite{Chung97}, Kirchhoff's theorem for the number of spanning
trees~\cite{Bollobas98}, and the dynamics of coupled
oscillators~\cite{BP02b,NMLH03}.  Its properties are the subject of
hundreds of papers in the mathematics and physics literature and are by now
quite well understood.  For our purposes, however, we will need only a few
simple observations about the matrix to make progress.

Our task is to choose the vector $\vec{s}$ so as to minimize the cut size,
Eq.~\eqref{defsr}.  Let us write $\vec{s}$ as a linear combination of the
normalized eigenvectors~$\vec{v}_i$ of the Laplacian thus:
$\vec{s}=\sum_{i=1}^n a_i \vec{v}_i$, where $a_i=\vec{v}_i^T\vec{s}$ and
the normalization $\vec{s}^T\vec{s}=n$ implies that
\begin{equation}
\sum_{i=1}^n a_i^2 = n.
\label{normalization}
\end{equation}
Then
\begin{equation}
R = \sum_i a_i \vec{v}_i^T \mat{L} \sum_j a_j \vec{v}_j
  = \sum_{ij} a_i a_j \lambda_j \delta_{ij}
  = \sum_i a_i^2 \lambda_i,
\label{rvalue}
\end{equation}
where $\lambda_i$ is the eigenvalue of $\mat{L}$ corresponding to the
eigenvector~$\vec{v}_i$ and we have made use of
$\vec{v}^T_i\vec{v}_j=\delta_{ij}$.  Without loss of generality, we assume
that the eigenvalues are labeled in increasing order
$\lambda_1\le\lambda_2\le\ldots\le\lambda_n$.  The task of minimizing~$R$
can then be equated with the task of choosing the nonnegative
quantities~$a_i^2$ so as to place as much as possible of the weight in the
sum~\eqref{rvalue} in the terms corresponding to the lowest eigenvalues,
and as little as possible in the terms corresponding to the highest, while
respecting the normalization constraint~\eqref{normalization}.

The sum of every row (and column) of the Laplacian matrix is zero:
\begin{equation}
\sum_j L_{ij} = \sum_j (k_i \delta_{ij} - A_{ij}) = k_i - k_i = 0,
\end{equation}
where we have made use of~\eqref{degree}.  Thus the vector $(1,1,1,\ldots)$
is always an eigenvector of the Laplacian with eigenvalue zero.  It is less
trivial, but still straightforward, to demonstrate that all eigenvalues of
the Laplacian are nonnegative.  (The Laplacian is symmetric and equal to
the square of the edge incidence matrix, and hence its eigenvalues are all
the squares of real vectors.)  Thus the eigenvalue~0 is always the smallest
eigenvalue of the Laplacian and the corresponding eigenvector is
$\vec{v}_1=(1,1,1,\ldots)/\sqrt{n}$, correctly normalized.

Given these observations it is now straightforward to see how to minimize
the cut size~$R$.  If we choose $\vec{s}=(1,1,1,\ldots)$, then all of the
weight in the final sum in Eq.~\eqref{rvalue} is in the term corresponding
to the lowest eigenvalue~$\lambda_1=0$ and all other terms are zero, since
$(1,1,1,\ldots)$ is an eigenvector and the eigenvectors are orthogonal.
Thus this choice gives us $R=0$, which is the smallest value it can take
since it is by definition a nonnegative quantity.

Unfortunately, when we consider the physical interpretation of this
solution, we see that it is trivial and uninteresting.  Given the
definition~\eqref{defss} of~$\vec{s}$, the choice $\vec{s}=(1,1,1,\ldots)$
is equivalent to placing all the vertices in group~1 and none of them in
group~2.  Technically, this is a valid division of the network, but it is
not a useful one.  Of course the cut size is zero if we put all the
vertices in one of the groups and none in the other, but such a trivial
solution tells us nothing about how to solve our original problem.

We would like to forbid this trivial solution, so as to force the method to
find a nontrivial one.  A variety of ways have been explored for achieving
this goal, of which the most common is to fix the sizes of the two groups,
which is convenient if, as discussed above, the sizes of the groups are
specified anyway as a part of the problem.  In the present case, fixing the
sizes of the groups fixes the coefficient~$a_1^2$ of the $\lambda_1$~term
in the sum in Eq.~\eqref{rvalue}; if the required sizes of the groups are
$n_1$ and~$n_2$, then
\begin{equation}
a_1^2 = \bigl( \vec{v}_1^T\vec{s} \bigr)^2 = {(n_1-n_2)^2\over n}.
\end{equation}
Since we cannot vary this coefficient, we shift our attention to the other
terms in the sum.  If there were no further constraints on our choice
of~$\vec{s}$, apart from the normalization condition~$\vec{s}^T\vec{s}=n$,
our course would be clear: $R$~would be minimized by choosing $\vec{s}$
proportional to the second eigenvector~$\vec{v}_2$ of the Laplacian, also
called the \defn{Fiedler vector}.  This choice places all of the weight in
Eq.~\eqref{rvalue} in the term involving the second-smallest
eigenvalue~$\lambda_2$, also known as the \defn{algebraic connectivity}.
The other terms would automatically be zero, since the eigenvectors are
orthogonal.

Unfortunately, there is an additional constraint on $\vec{s}$ imposed by
the condition, Eq.~\eqref{defss}, that its elements take the values~$\pm1$,
which means in most cases that $\vec{s}$ cannot be chosen parallel
to~$\vec{v}_2$.  This makes the optimization problem much more difficult.
Often, however, quite good approximate solutions can be obtained by
choosing $\vec{s}$ to be as close to parallel with $\vec{v}_2$ as possible.
This means maximizing the quantity
\begin{equation}
\bigl| \vec{v}_2^T\vec{s} \bigr| = \biggl| \sum_i v^{(2)}_i s_i \biggr|
  \le \sum_i \bigl| v^{(2)}_i \bigr|,
\label{tomax}
\end{equation}
where $v^{(2)}_i$ is the $i$th element of $\vec{v}_2$.  Here the second
relation follows via the triangle inequality, and becomes an equality only
when all terms in the first sum are positive (or negative).  In other
words, the maximum of $|\vec{v}_2^T\vec{s}|$ is achieved when $v^{(2)}_i
s_i\ge0$ for all~$i$, or equivalently when $s_i$ has the same sign as
$v^{(2)}_i$.  Thus the maximum is obtained with the choice
\begin{equation}
s_i = \begin{cases}
        +1 & \quad\text{if $v^{(2)}_i\ge0$,} \\
        -1 & \quad\text{if $v^{(2)}_i<0$.}
      \end{cases}
\end{equation}
Even this choice however is often forbidden by the condition that the
number of $+1$ and $-1$ elements of $\vec{s}$ be equal to the desired sizes
$n_1$ and $n_2$ of the two groups, in which case the best solution is
achieved by assigning vertices to one of the groups in order of the
elements in the Fiedler vector, from most positive to most negative, until
the groups have the required sizes.  For groups of different sizes there
are two distinct ways of doing this, one in which the smaller group
corresponds to the most positive elements of the vector and one in which
the larger group does.  We can choose between them by calculating the cut
size~$R$ for both cases and keeping the one that gives the better result.

\begin{figure}
\begin{center}
\includegraphics[width=\columnwidth]{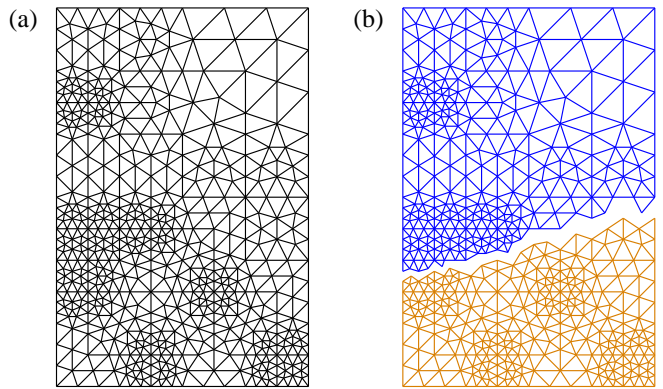}
\end{center}
\caption{(a)~The mesh network of Bern~\etal~\cite{BEG90}.  (b)~The best
division into equal-sized parts found by the spectral partitioning
algorithm based on the Laplacian matrix.}
\label{eppstein}
\end{figure}

This then is the spectral partitioning method in its simplest form.  It is
not guaranteed to minimize~$R$, but, particularly in cases where
$\lambda_2$ is well separated from the eigenvalues above it, it often does
very well.  Figure~\ref{eppstein} shows an example application typical of
those found in the literature, to a two-dimensional mesh such as might be
used in parallel finite-element calculations.  This particular mesh is a
small 547-vertex example from Bern~\etal~\cite{BEG90} and is shown complete
in panel~(a) of the figure.  Panel~(b) shows the division of the mesh into
two parts of $273$ and $274$ vertices respectively using the spectral
partitioning approach, which finds a cut of size 46 edges in this case.

Although the cut found in this example is a reasonable one, it does not
appear---at least to this author's eye---that the vertex groups in
Fig.~\ref{eppstein}b constitute any kind of natural division of the network
into ``communities.''  This is typical of the problems to which spectral
partitioning is usually applied: in most circumstances the network in
question does not divide up easily into groups of the desired sizes, but
one must do the best one can.  For these types of tasks, spectral
partitioning is an effective and appropriate tool.  The task of finding
natural community divisions in a network, however, is quite different, and
demands a different approach, as we now discuss.

\section{Community structure and modularity}
\label{modsec}
Despite its evident success in the graph partitioning arena, spectral
partitioning is a poor approach for detecting natural community structure
in real-world networks, which is the primary topic of this paper.  The
issue is with the condition that the sizes of the groups into which the
network is divided be fixed.  This condition is neither appropriate nor
realistic for community detection problems.  In most cases we do not know
in advance the sizes of the communities in a network and choosing arbitrary
sizes will usually preclude us from finding the best solution to the
problem.  We would like instead to let the group sizes be free, but the
spectral partitioning method breaks down if we do this, as we have seen: if
the group sizes are not fixed, then the minimum cut size is always achieved
by putting all vertices in one group and none in the other.  Indeed, this
statement is considerably broader than the spectral partitioning method
itself, since any method that correctly minimizes the cut size without
constraint on the group sizes is sure to find, in the general case, that
the minimum value is achieved for this same trivial division.

Several approaches have been proposed to get around this problem.  For
instance, the \defn{ratio cut} method~\cite{WC89} minimizes not the simple
cut size~$R$ but the ratio $R/(n_1 n_2)$, where $n_1$ and $n_2$ are again
the sizes of the two groups of vertices.  This penalizes configurations in
which either of the groups is small and hence favors balanced divisions
over unbalanced ones, releasing us from the obligation to fix the group
sizes.  Spectral algorithms based on ratio cuts have been
proposed~\cite{HK92,CSZ93} and have proved useful for certain classes of
partitioning problems.  Still, however, this approach effectively chooses
the group sizes, at least approximately, since it is biased in favor of
divisions into equal-sized parts.  Variations are possible that are biased
towards other, unequal part sizes, but then one must choose those parts
sizes and so again we have a situation in which we need to know in advance
the sizes of the groups if we are to get the ``right'' results.  The ratio
cut method does allow some leeway for the sizes to vary around their
specified values, which makes it more flexible than the simple minimum cut
method, but at its core it still suffers from the same drawbacks that make
standard spectral partitioning inappropriate for community detection.

The fundamental problem with all of these methods is that cut sizes are
simply not the right thing to optimize because they don't accurately
reflect our intuitive concept of network communities.  A good division of a
network into communities is not merely one in which the number of edges
running between groups is small.  Rather, it is one in which the number of
edges between groups is \emph{smaller than expected}.  Only if the number
of between-group edges is significantly lower than would be expected purely
by chance can we justifiably claim to have found significant community
structure.  Equivalently, we can examine the number of edges \emph{within}
communities and look for divisions of the network in which this number is
higher than expected---the two approaches are equivalent since the total
number of edges is fixed and any edges that do not lie between communities
must necessarily lie inside them.

These considerations lead us to shift our attention from measures based on
pure cut size to a modified benefit function~$Q$ defined by
\begin{eqnarray}
Q &=& \mbox{(number of edges within communities)} \nonumber\\
  & & \quad{} - \mbox{(expected number of such edges)}.
\label{defsq1}
\end{eqnarray}
This benefit function is called \defn{modularity}~\cite{Newman03c,NG04}.
It is a function of the particular division of the network into groups,
with larger values indicating stronger community structure.  Hence we
should, in principle, be able to find good divisions of a network into
communities by optimizing the modularity over possible divisions.  This
approach, proposed in~\cite{Newman04a} and since pursued by a number of
authors~\cite{CNM04,GSA04,GA05,DA05,Newman06b}, has proven highly effective
in practice~\cite{DDDA05} and is the primary focus of this article.

The first term in Eq.~\eqref{defsq1} is straightforward to calculate.  The
second, however, is rather vague and needs to be made more precise before
we can evaluate the modularity.  What exactly do we mean by the ``expected
number'' of edges within a community?  Answering this question is
essentially equivalent to choosing a ``null model'' against which to
compare our network.  The definition of the modularity involves a
comparison of the number of within-group edges in a real network and the
number in some equivalent randomized model network in which edges are
placed without regard to community structure.

It is one of the strengths of the modularity approach that it makes the
role of this null model explicit and clear.  All methods for finding
communities are, in a sense, assuming some null model, since any method
must make a value judgment about when a particular density of edges is
significant enough to define a community.  In most cases, this assumption
is hidden within the workings of a computer algorithm and is difficult to
disentangle, even when the algorithm itself is well understood.  By
bringing its assumptions out into the open, the modularity method gives us
more control over our calculations and more understanding of their
implications.

Our null model must have the same number of vertices~$n$ as the original
network, so that we can divide it into the same groups for comparison, but
apart from this we have a good deal of freedom about our choice of model.
We here consider the broad class of randomized models in which we specify
separately the probability~$P_{ij}$ for an edge to fall between every pair
of vertices~$i,j$.  More precisely, $P_{ij}$~is the expected number of
edges between $i$ and~$j$, a definition that allows for the possibility
that there may be more than one edge between a pair of vertices, which
happens in certain types of networks.  We will consider some particular
choices of $P_{ij}$ in a moment, but for now let us pursue the developments
in general form.

Given $P_{ij}$, the modularity can be defined as follows.  The actual
number of edges falling between a particular pair of vertices $i$ and $j$
is~$A_{ij}$, Eq.~\eqref{adjacency}, and the expected number is, by
definition,~$P_{ij}$.  Thus the actual minus expected number of edges
between $i$ and $j$ is $A_{ij}-P_{ij}$ and the modularity is (proportional
to) the sum of this quantity over all pairs of vertices belonging to the
same community.  Let us define $g_i$ to be the community to which
vertex~$i$ belongs.  Then the modularity can be written
\begin{equation}
Q = {1\over2m} \sum_{ij} \bigl[ A_{ij} - P_{ij} \bigr] \delta(g_i,g_j),
\label{q1}
\end{equation}
where $\delta(r,s)=1$ if $r=s$ and 0 otherwise and $m$ is again the number
of edges in the network.  The extra factor of $1/2m$ in Eq.~\eqref{q1} is
purely conventional; it is included for compatibility with previous
definitions of the modularity and plays no part in the maximization of~$Q$
since it is a constant for any given network.  A special case of
Eq.~\eqref{q1} was given previously by the present author
in~\cite{Newman04f} and independently, in slightly different form, by White
and Smyth~\cite{WS05}.  A number of other expressions for the modularity
have also been presented by various authors~\cite{NG04,GSA04,DA05} and are
convenient in particular applications.  Also of interest is the derivation
of the modularity given recently by Reichardt and Bornholdt~\cite{RB06},
which is quite general and provides an interesting alternative to the
derivation presented here.

Returning to the null model, how should $P_{ij}$ be chosen?  The choice is
not entirely unconstrained.  First, we consider in this paper only
undirected networks, which implies that $P_{ij}=P_{ji}$.  Second, it is
axiomatically the case that $Q=0$ when all vertices are placed in a single
group together: by definition, the number of edges within groups and the
expected number of such edges are both equal to~$m$ in this case.  Setting
all $g_i$ equal in Eq.~\eqref{q1}, we find that $\sum_{ij} [ A_{ij} -
P_{ij} ] = 0$ or equivalently
\begin{equation}
\sum_{ij} P_{ij} = \sum_{ij} A_{ij} = 2m.
\label{sanity}
\end{equation}
This equation says that we are restricted to null models in which the
expected number of edges in the entire network equals the actual number of
edges in the original network---a natural choice if our comparison of
numbers of edges within groups is to have any meaning.

Beyond these basic considerations, there are many possible choices of null
model and several have been considered previously in the
literature~\cite{NG04,RB04,MD05}.  Perhaps the simplest is the standard
(Bernoulli) random graph, in which edges appear with equal probability
$P_{ij}=p$ between all vertex pairs.  With a suitably chosen value of $p$
this model can be made to satisfy~\eqref{sanity} but, as many authors have
pointed out~\cite{Strogatz01,DM02,WS98}, the model is not a good
representation of most real-world networks.  A particularly glaring aspect
in which it errs is its degree distribution.  The random graph has a
binomial degree distribution (or Poisson in the limit of large graph size),
which is entirely unlike the right-skewed degree distributions found in
most real-world networks~\cite{BA99b,ASBS00}.  A much better null model
would be one in which the degree distribution is approximately the same as
that of the real-world network of interest.  To satisfy this demand we will
restrict our attention in this paper to models in which the expected degree
of each vertex within the model is equal to the actual degree of the
corresponding vertex in the real network.  Noting that the expected degree
of vertex~$i$ is given by $\sum_j P_{ij}$, we can express this condition as
\begin{equation}
\sum_j P_{ij} = k_i.
\label{constraint}
\end{equation}
If this constraint is satisfied, then~\eqref{sanity} is automatically
satisfied as well, since $\sum_i k_i = 2m$.

Equation~\eqref{constraint} is a considerably more stringent constraint
than~\eqref{sanity}---in most cases, for instance, it excludes the
Bernoulli random graph---but it is one that we believe makes good sense,
and one moreover that has a variety of desirable consequences for the
developments that follow.

The simplest null model in this class, and the only one that has been
considered at any length in the past, is the model in which edges are
placed entirely at random, subject to the constraint~\eqref{constraint}.
That is, the probability that an end of a randomly chosen edge attaches to
a particular vertex~$i$ depends only on the expected degree~$k_i$ of that
vertex, and the probabilities for the two ends of a single edge are
independent of one another.  This implies that the expected number of edges
$P_{ij}$ between vertices~$i$ and $j$ is the product $f(k_i)f(k_j)$ of
separate functions of the two degrees, where the functions must be the same
since $P_{ij}$ is symmetric.  Then Eq.~\eqref{constraint} implies
\begin{equation}
\sum_{j=1}^n P_{ij} = f(k_i) \sum_{j=1}^n f(k_j) = k_i,
\end{equation}
for all~$i$ and hence $f(k_i)=Ck_i$ for some constant~$C$.  And
Eq.~\eqref{sanity} says that
\begin{equation}
2m = \sum_{ij} P_{ij} = C^2 \sum_{ij} k_i k_j = (2mC)^2,
\end{equation}
and hence $C=1/\sqrt{2m}$ and
\begin{equation}
P_{ij} = {k_ik_j\over2m}.
\label{defscm}
\end{equation}
This model has been studied in the past in its own right as a model of a
network, for instance by Chung and Lu~\cite{CL02a}.  It is also closely
related to the \defn{configuration model}, which has been studied widely in
the mathematics and physics literature~\cite{Luczak92,MR95,NSW01,CL02a}.
Indeed, essentially all expected properties of our model and the
configuration model are identical in the limit of large network size, and
hence Eq.~\eqref{defscm} can be considered equivalent to the configuration
model in this limit.\footnote{The technical difference between the two
models is that the configuration model is a random multigraph conditioned
on the actual degree sequence, while the model used here is a random
multigraph conditioned on the expected degree sequence.  This makes the
ensemble of the former considerably smaller than that of the latter, but
the difference is analogous to the difference between canonical and grand
canonical ensembles in statistical mechanics and the two give the same
answers in the thermodynamic limit for roughly the same reason.  In
particular, we note that the probability of an edge falling between two
vertices $i$ and $j$ in the configuration model is also given by
Eq.~\eqref{defscm} in the limit of large network size; for smaller
networks, there are corrections of order~$1/n$.}

Although many of the developments outlined in this paper are true for quite
general choices of the null model used to define the modularity, the
choice~\eqref{defscm} is the only one we will pursue here.  It is worth
keeping mind however that other choices are possible: Massen and
Doye~\cite{MD05}, for instance, have used a variant of the configuration
model in which multiedges and self-edges were excluded.  And further
choices could be useful in specific cases, such as cases where there are
strong correlations between the degrees of vertices~\cite{PVV01,Newman02f}
or where there is a high level of network transitivity~\cite{WS98}.

\section{Spectral optimization of modularity}
\label{specmod}
Once we have an explicit expression for the modularity we can determine the
community structure by maximizing it over possible divisions of the
network.  Unfortunately, exhaustive maximization over all possible
divisions is computational intractable because there are simply too many
divisions, but various approximate optimization methods have proven
effective~\cite{Newman04a,CNM04,GSA04,GA05,MD05,RB06,DA05}.  Here, we
develop a matrix-based approach analogous to the spectral partitioning
method of Section~\ref{specpart}, which leads not only to a whole array of
possible optimization algorithms but also to new insights about the nature
and implications of community structure in networks.

\subsection{Leading eigenvector method}
\label{singlevec}
As before, let us consider initially the division of a network into just
two communities and denote a potential such division by an index
vector~$\vec{s}$ with elements as in Eq.~\eqref{defss}.  We notice that the
quantity $\half(s_is_j+1)$ is 1 if $i$ and $j$ belong to the same group and
0 if they belong to different groups or, in the notation of Eq.~\eqref{q1},
\begin{equation}
\delta(g_i,g_j) = \half (s_is_j+1).
\end{equation}
Thus we can write~\eqref{q1} in the form
\begin{eqnarray}
Q &=& {1\over4m} \sum_{ij} \bigl[ A_{ij} - P_{ij} \bigr] (s_is_j+1)
      \nonumber\\
  &=& {1\over4m} \sum_{ij} \bigl[ A_{ij} - P_{ij} \bigr] s_is_j,
\end{eqnarray}
where we have in the second line made use of Eq.~\eqref{sanity}.  This
result can conveniently be rewritten in matrix form as
\begin{equation}
Q = {1\over4m} \vec{s}^T \mat{B} \vec{s},
\label{q2}
\end{equation}
where $\mat{B}$ is the real symmetric matrix having elements
\begin{equation}
B_{ij} = A_{ij} - P_{ij}.
\label{defsbij}
\end{equation}
We call this matrix the \defn{modularity matrix} and it plays a role in the
maximization of the modularity equivalent to that played by the Laplacian
in standard spectral partitioning: Equation~\eqref{q2} is the equivalent of
Eq.~\eqref{defsr} for the cut size and matrix methods can thus be applied
to the modularity that are the direct equivalents of those developed for
spectral partitioning, as we now show.

First, let us point out a few important properties of the modularity
matrix.  Equations~\eqref{degree} and~\eqref{constraint} together imply
that all rows (and columns) of the modularity matrix sum to zero:
\begin{equation}
\sum_j B_{ij} = \sum_j A_{ij} - \sum_j P_{ij}
              = k_i - k_i = 0.
\end{equation}
This immediately implies that for any network the vector $(1,1,1,\ldots)$
is an eigenvector of the modularity matrix with eigenvalue zero, just as is
the case with the Laplacian.  Unlike the Laplacian however, the eigenvalues
of the modularity matrix are not necessarily all of one sign and in
practice the matrix usually has both positive and negative eigenvalues.
This observation---and the eigenspectrum of the modularity matrix in
general---are, as we will see, closely tied to the community structure of
the network.

Working from Eq.~\eqref{q2} we now proceed by direct analogy with
Section~\ref{specpart}.  We write $\vec{s}$ as a linear combination of the
normalized eigenvectors~$\vec{u}_i$ of the modularity matrix, $\vec{s} =
\sum_{i=1}^n a_i \vec{u}_i$ with $a_i=\vec{u}_i^T\vec{s}$.  Then
\begin{equation}
Q = {1\over4m} \sum_i a_i^2 \beta_i,
\label{q3}
\end{equation}
where $\beta_i$ is the eigenvalue of $\mat{B}$ corresponding to the
eigenvector~$\vec{u}_i$.  We now assume that the eigenvalues are labeled in
\emph{decreasing} order $\beta_1\ge\beta_2\ge\ldots\ge\beta_n$ and the task
of maximizing~$Q$ is one of choosing the quantities~$a_i^2$ so as to place
as much as possible of the weight in the sum~\eqref{q3} in the terms
corresponding to the largest (most positive) eigenvalues.

As with ordinary spectral partitioning, this would be a simple task if our
choice of $\vec{s}$ were unconstrained (apart from normalization): we would
just choose $\vec{s}$ proportional to the leading eigenvector $\vec{u}_1$
of the modularity matrix.  But the elements of $\vec{s}$ are restricted to
the values $s_i=\pm1$, which means that $\vec{s}$ cannot normally be chosen
parallel to~$\vec{u}_1$.  Again as before, however, good approximate
solutions can be obtained by choosing $\vec{s}$ to be as close to parallel
with $\vec{u}_1$ as possible, which is achieved by setting
\begin{equation}
s_i = \begin{cases}
        +1 & \quad\text{if $u^{(1)}_i\ge0$,} \\
        -1 & \quad\text{if $u^{(1)}_i<0$.}
      \end{cases}
\end{equation}
This then is our first and simplest algorithm for community detection: we
find the eigenvector corresponding to the most positive eigenvalue of the
modularity matrix and divide the network into two groups according to the
signs of the elements of this vector.

\begin{figure}
\begin{center}
\includegraphics[width=\columnwidth]{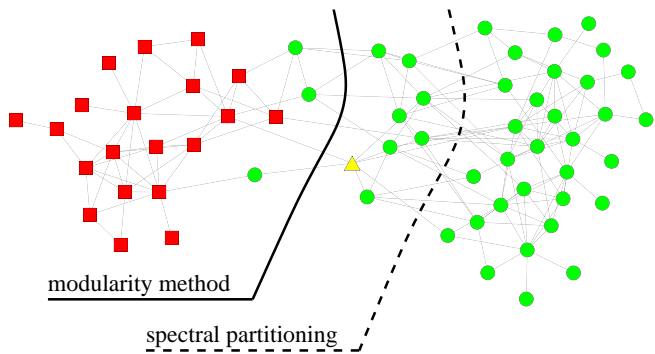}
\end{center}
\caption{The dolphin social network of Lusseau~\etal~\cite{Lusseau03a}.
The dashed curve represents the division into two equally sized parts found
by a standard spectral partitioning calculation (Section~\ref{specpart}).
The solid curve represents the division found by the modularity-based
method of this section.  And the squares and circles represent the actual
division of the network observed when the dolphin community split into two
as a result of the departure of a keystone individual.  (The individual who
departed is represented by the triangle.)}
\label{dolphins}
\end{figure}

\begin{figure*}
\begin{center}
\includegraphics[width=12cm]{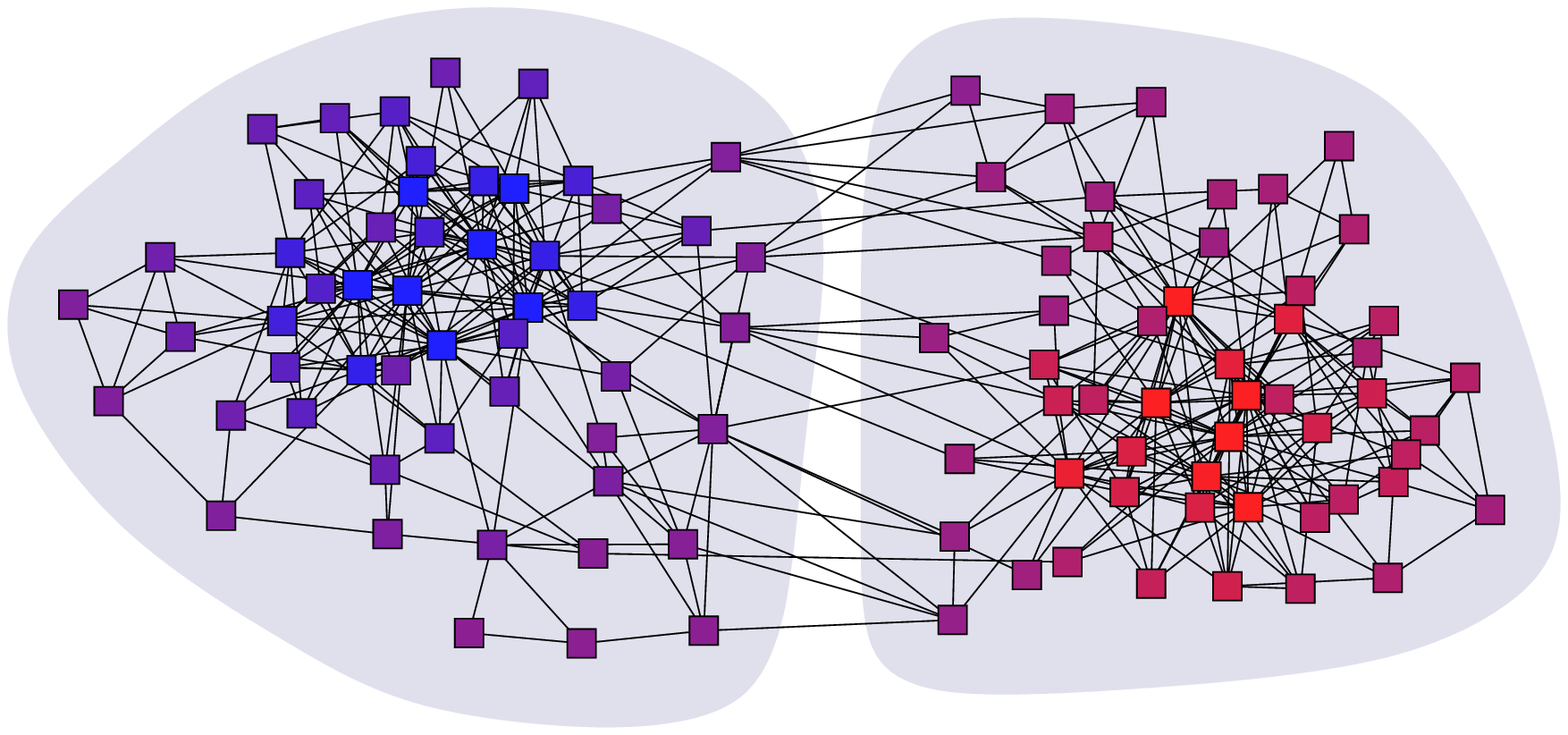}
\end{center}
\caption{The network of political books described in the text.  Vertex
colors range from blue to red to represent the values of the corresponding
elements of the leading eigenvector of the modularity matrix.}
\label{books}
\end{figure*}

In practice, this method works nicely, as discussed in~\cite{Newman06b}.
Making the choice~\eqref{defscm} for our null model, we have applied it to
a variety of standard and less standard test networks and find that it does
a good job of finding community divisions.  Figure~\ref{dolphins} shows a
representative example, an animal social network assembled and studied by
Lusseau~\etal~\cite{Lusseau03a}.  The vertices in this network represent 62
bottlenose dolphins living in Doubtful Sound, New Zealand, with social ties
between dolphin pairs established by direct observation over a period of
several years.  This network is of particular interest because, during the
course of the study, the dolphin group split into two smaller subgroups
following the departure of a key member of the population.  The subgroups
are represented by the shapes of the vertices in the figure.  The dotted
line denotes the division of the network into two equal-sized groups found
by the standard spectral partitioning method.  While, as expected, this
method does a creditable job of dividing the network into groups of these
particular sizes, it is clear to the eye that this is not the natural
community division of the network and neither does it correspond to the
division observed in real life.  The spectral partitioning method is
hamstrung by the requirement that we specify the sizes of the two
communities; unless we know what they are in advance, blind application of
the method will not usually find the ``right'' division of the network.

The method based on the leading eigenvector of the modularity matrix,
however, does much better.  Unconstrained by the need to find groups of any
particular size, this method finds the division denoted by the solid line
in the figure, which, as we see, corresponds quite closely to the split
actually observed---all but three of the 62 dolphins are placed in the
correct groups.

The magnitudes of the elements of the eigenvector $\vec{u}_1$ also contain
useful information about the network, indicating, as discussed
in~\cite{Newman06b}, the ``strength'' with which vertices belong to the
communities in which they are placed.  As an example of this phenomenon
consider Fig.~\ref{books}, which depicts the network of political books
from Ref.~\cite{Newman06b}.  This network, compiled by V.~Krebs
(unpublished), represents recent books on US politics, with edges
connecting pairs of books that are frequently purchased by the same
customers of the on-line bookseller Amazon.com.  Applying our method, we
find that the network divides as shown in the figure, with the colors of
the vertices representing the values of the elements of the eigenvector.
The two groups correspond closely to the apparent alignment of the books
according to left-wing and right-wing points of view~\cite{Newman06b}, and
are suggestively colored blue and red in the figure.\footnote{By a fluke of
recent history, the colors blue and red have come to denote liberal and
conservative points of view respectively in US politics, where in most
other parts of the world the color-scheme is the other way around.}  The
most blue and most red vertices are those that, by our calculation, belong
most strongly to the two groups and are thus, perhaps, the ``most
left-wing'' and ``most right-wing'' of the books under consideration.
Those familiar with current US politics will be unsurprised to learn that
the most left-wing book in this sense was the polemical \emph{Bushwacked}
by Molly Ivins and Lou Dubose.  Perhaps more surprising is the most
right-wing book: \emph{A National Party No More} by Zell
Miller.\footnote{Miller is a former Democratic (i.e.,~ostensibly liberal)
governor and US senator for the state of Georgia.  He became known in the
later years of his career, however, for views that aligned more closely
with the conservative Republicans than with the Democrats.  Even so, Miller
was never the most conservative member of the senate, nor is his book the
most conservative in this study.  But our measure is not based on the
content of the books; it merely finds the vertices in the network that are
most central to their communities.  The ranking of Miller's book in this
calculation results from its centrality within the community of
conservative book buying.  This book, while not in fact as right-wing as
some, apparently appeals widely and exclusively to conservatives,
presumably because of the unusual standing of its author as a nominal
Democrat supporting the Republican cause.}

\subsection{Other eigenvectors of the modularity matrix}
\label{multivec}
The algorithm described in the previous section has two obvious
shortcomings.  First, it divides networks into only two communities, while
real-world networks can certainly have more than two.  Second, it makes use
only of the leading eigenvector of the modularity matrix and ignores all
the others, which throws away useful information contained in those other
vectors.  Both of these shortcomings are remedied by the following
generalization of the method.

Consider the division of a network into~$c$ non-overlapping communities,
where $c$ may now be greater than~2.  Following Alpert and Yao~\cite{AY95}
and more recently White and Smyth~\cite{WS05}, let us define an $n\times c$
index matrix $\mat{S}$ with one column for each community:
$\mat{S}=(\vec{s}_1|\vec{s}_2|\ldots|\vec{s}_c)$.  Each column is an index
vector now of $(0,1)$ elements (rather than $\pm1$ as previously), such
that
\begin{equation}
S_{ij} = \begin{cases}
           1 & \quad\text{if vertex~$i$ belongs to community~$j$,}\\
           0 & \quad\text{otherwise.}
         \end{cases}
\end{equation}
Note that the columns of $\mat{S}$ are mutually orthogonal, that the rows
each sum to unity, and that the matrix satisfies the normalization
condition $\Tr(\mat{S}^T\mat{S})=n$.

Observing that the $\delta$-symbol in Eq.~\eqref{q1} is now given by
\begin{equation}
\delta(g_i,g_j) = \sum_{k=1}^c S_{ik} S_{jk},
\end{equation}
the modularity for this division of the network is
\begin{equation}
Q = \sum_{i,j=1}^n\:\sum_{k=1}^c B_{ij} S_{ik} S_{jk}
  = \Tr(\mat{S}^T\mat{B}\mat{S}),
\label{matq}
\end{equation}
where here and henceforth we suppress the leading multiplicative
constant~$1/2m$ from Eq.~\eqref{q1}, which has no effect on the position of
the maximum of the modularity.

Writing $\mat{B}=\mat{U}\mat{D}\mat{U}^T$, where
$\mat{U}=(\vec{u}_1|\vec{u}_2|\ldots)$ is the matrix of eigenvectors
of~$\mat{B}$ and $\mat{D}$ is the diagonal matrix of eigenvalues
$D_{ii}=\beta_i$, we then find that
\begin{equation}
Q = \sum_{j=1}^n \sum_{k=1}^c \beta_j (\vec{u}_j^T\vec{s}_k)^2.
\label{qmulti}
\end{equation}
Again we wish to maximize this modularity, but now we have no constraint on
the number~$c$ of communities; we can give~$\mat{S}$ as many columns as we
like in our effort to make~$Q$ as large as possible.

If the elements of the matrix $\mat{S}$ were unconstrained apart from the
basic conditions on the rows and columns mentioned above, a choice of $c$
communities would be equivalent to choosing $c-1$ independent, mutually
orthogonal columns $\vec{s}_1\ldots\vec{s}_{c-1}$.  (Only $c-1$ of the
columns are independent, the last being fixed by the condition that the
rows of $\mat{S}$ sum to unity.)  In this case our path would be clear:
$Q$~would be maximized by choosing the columns proportional to the leading
eigenvectors of~$\mat{B}$.  However, only those eigenvectors corresponding
to positive eigenvalues can give positive contributions to the modularity,
so the optimal modularity would be achieved by choosing exactly as many
independent columns of $\mat{S}$ as there are positive eigenvalues, or
equivalently by choosing the number of groups~$c$ to be 1 greater than the
number of positive eigenvalues.

Unfortunately, our problem has the additional constraint that the index
vectors~$\vec{s}_i$ have only binary $(0,1)$ elements, which means it may
not be possible to find as many index vectors making positive contributions
to the modularity as the set of positive eigenvalues suggests.  Thus the
number of positive eigenvalues, plus~1, is an \emph{upper bound} on the
number of communities and again we see that there is an intimate connection
between the properties of the modularity matrix and the community structure
of the network it describes.

\subsection{Vector partitioning algorithm}
\label{vectorpart}
In Section~\ref{singlevec} we maximized the modularity approximately by
focusing solely on the term in $Q$ proportional to the largest eigenvalue
of~$\mat{B}$.  Let us now make the more general (and often better)
approximation of keeping the leading $p$ eigenvalues, where $p$ may be
anywhere between 1 and~$n$.  Some of the eigenvalues, however, may be
negative, which will prove inconvenient.  To get around this we rewrite
Eq.~\eqref{matq} thus:
\begin{eqnarray}
Q &=& n\alpha + \Tr[\mat{S}^T\mat{U}(\mat{D}-\alpha\mat{I})\mat{U}^T\mat{S}]
      \nonumber\\
  &=& n\alpha + \sum_{j=1}^n \sum_{k=1}^c (\beta_j-\alpha)
      \biggl[ \sum_{i=1}^n U_{ij} S_{ik} \biggr]^2,
\label{alphaq}
\end{eqnarray}
where $\alpha$ is a constant whose value we will choose shortly and we have
made use of $\Tr(\mat{S}^T\mat{S})=n$ and the fact that $\mat{U}$ is
orthogonal.

Now, employing an argument similar to that used for ordinary spectral
partitioning in~\cite{AY95}, let us define a set of \defn{vertex
vectors}~$\vec{r}_i$, $i=1\ldots n$, of dimension~$p$, such that the $j$th
component of the $i$th vector is
\begin{equation}
\bigl[ \vec{r}_i \bigr]_j = \sqrt{\beta_j-\alpha}\,U_{ij}.
\label{defsri}
\end{equation}
Provided we choose $\alpha\le\beta_p$, $\vec{r}_i$~is guaranteed real for
all~$i$.  Then, dropping terms in~\eqref{alphaq} proportional to the
smallest $n-p$ of the factors~$\beta_j-\alpha$, we have
\begin{eqnarray}
Q &\simeq& n\alpha + \sum_{j=1}^p \sum_{k=1}^c
      \biggl[ \sum_{i=1}^n \sqrt{\beta_j-\alpha}\,U_{ij} 
      S_{ik} \biggr]^2 \nonumber\\
  &=& n\alpha + \sum_{k=1}^c \sum_{j=1}^p
      \biggl[ \sum_{i\in G_k} \bigl[ \vec{r}_i \bigr]_j \biggr]^2 \nonumber\\
  &=& n\alpha + \sum_{k=1}^c | \vec{R}_k |^2,\qquad{}
\label{finalq}
\end{eqnarray}
where $G_k$ is the set of vertices comprising group~$k$ and the
\defn{community vectors}~$\vec{R}_k$, $k=1\ldots c$, are
\begin{equation}
\vec{R}_k = \sum_{i\in G_k} \vec{r}_i.
\label{defsrk}
\end{equation}

The community structure problem is now equivalent to choosing a division of
the vertices into groups so as to maximize the magnitudes of the
vectors~$\vec{R}_k$.  This means we need to arrange that the individual
vertex vectors~$\vec{r}_i$ going into each group point in approximately the
same direction.  Problems of this type are called \defn{vector
partitioning} problems.

The parameter~$p$ controls the balance between the complexity of the vector
partitioning problem and the accuracy of the approximation we make by
keeping only some of the eigenvalues.  The calculations will be faster but
less accurate for smaller~$p$ and slower but more accurate for larger.  For
the special case $p=n$ where we keep all of the eigenvalues,
Eq.~\eqref{finalq} is exact.  In this case, we note that the vertex vectors
have the property
\begin{equation}
\vec{r}_i^T\vec{r}_j = \sum_{k=1}^n U_{ik} (\beta_k-\alpha) U_{jk}
  = B_{ij} - \alpha\delta_{ij}.
\label{rirj}
\end{equation}
It's then simple to see that Eq.~\eqref{finalq} is trivially equivalent to
the fundamental definition~\eqref{q1} of the modularity, so in the $p=n$
case our mapping to a vector partitioning problem gives little insight into
the modularity maximization problem.  The real advantage of our approach
comes when $p<n$, where the method extracts precisely those factors that
make the principal contributions to the modularity---i.e.,~those
corresponding to the largest eigenvalues---discarding those that have
relatively little effect.  In practice, as we have seen for the
single-eigenvector algorithm, the main features of the community structure
are often captured by just the first eigenvector or perhaps the first few,
which allows us to reduce the complexity of our optimization problem
immensely.

The approach is similar in concept to the standard technique of principal
components analysis (PCA) used to reduce high-dimensional data sets to
manageably small dimension by focusing on the eigendirections along which
the variance about the mean is greatest and ignoring directions that
contribute little.  In fact, this similarity is more than skin-deep: the
form of our modularity matrix is closely analogous to the covariance matrix
whose eigenvectors are the basis for PCA.  The elements of the covariance
matrix are correlation functions of the form $\av{xy}-\av{x}\av{y}$, where
$x$ and $y$ denote measured variables in the data set.  Thus the covariance
is the difference between the actual value of the mean product $\av{xy}$ of
two variables and the value $\av{x}\av{y}$ expected by chance for that
product if the variables were uncorrelated.  Similarly, the elements
$B_{ij}=A_{ij}-k_ik_j/2m$ of the modularity matrix are equal to the actual
number of edges $A_{ij}$ between a given pair of vertices minus the number
$k_ik_j/2m$ expected by chance, expressed in a product form.  In a sense,
our spectral method for modularity optimization can be regarded as a
``principal components analysis for networks.''  This aspect of the method
is clear, for instance, in the study of political books represented in
Fig.~\ref{books}: the leading eigenvector used to assign the colors to the
vertices in the figure is playing a role equivalent to the eigendirections
in PCA, defining a ``direction of greatest variation'' in the structure of
the network.  The vertex vectors of Eq.~\eqref{defsri} are similarly
analogous to the low-dimensional projections used in PCA.\footnote{This
suggests, for instance, that the vertex vectors for $p=2$ or~3 could be
used to define graph layouts for visualizing networks in 2 or 3 dimensions.
Either the endpoints of the vectors could define vertex positions
themselves or they could be used as starting positions for a spring
embedding visualizer or other more conventional layout scheme.}

Returning to our algorithm, let us consider again the special case of the
division of a network into just two communities.  (Multi-way division is
considered in Section~\ref{multiway}.)  Since $(1,1,1,\ldots)$ is always an
eigenvector of the modularity matrix and the eigenvectors are orthogonal,
the elements of all other eigenvectors must sum to zero:
\begin{equation}
\sum_{i=1}^n \bigl[ \vec{u}_j \bigr]_i
  = \sqrt{n}\,\vec{u}_1^T\vec{u}_j = 0.
\end{equation}
But Eq.~\eqref{defsri} then implies that
\begin{equation}
\sum_{i=1}^n \bigl[ \vec{r}_i \bigr]_j
  = \sqrt{\beta_j-\alpha} \sum_{i=1}^n U_{ij} = \sqrt{\beta_j-\alpha}
  \sum_{i=1}^n \bigl[ \vec{u}_j \bigr]_i = 0,
\end{equation}
and hence
\begin{equation}
\sum_{i=1}^n \vec{r}_i = 0
\end{equation}
for any value of~$p$.  This in turn implies that the community vectors
$\vec{R}_k$ also sum to zero:
\begin{equation}
\sum_{k=1}^c \vec{R}_k
  = \sum_{k=1}^c \:\sum_{i\in G_k} \vec{r}_i
  = \sum_{i=1}^n \vec{r}_i = 0.
\end{equation}
And as a special case of this last result, any division of a network into
two communities has community vectors $\vec{R}_1$ and $\vec{R}_2$ that are
of equal magnitude and oppositely directed.

Furthermore, the maximum of the modularity, Eq.~\eqref{finalq}, is always
achieved when each individual vertex vector~$\vec{r}_i$ has a positive
inner product with the community vector of the community to which the
vertex belongs.  To see this, observe that removing a vertex~$i$ from a
community~$k$ where $\vec{R}_k\cdot\vec{r}_i<0$ produces a change in the
corresponding term $|\vec{R}_k|^2$ in Eq.~\eqref{finalq} of
\begin{equation}
|\vec{R}_k-\vec{r}_i|^2 - |\vec{R}_k|^2
  = |\vec{r}_i|^2 - 2\vec{R}_k\cdot\vec{r}_i > 0.
\end{equation}
Similarly adding vertex~$i$ to a community for which
$\vec{R}_k\cdot\vec{r}_i>0$ also increases~$|\vec{R}_k|^2$.  Hence, we can
always increase the modularity by moving vertices until they are in groups
such that $\vec{R}_k\cdot\vec{r}_i>0$.

Taken together, these results imply that possible candidates for the
optimal division of a network into two groups are fully specified by just
the \emph{direction} of the single vector~$\vec{R}_1$.  Once we have this
direction, we know that the vertices divide according to whether their
projection along this direction is positive or negative.  Alternatively, we
can consider the direction of $\vec{R}_1$ to define a perpendicular plane
through the origin in the $p$-dimensional vector space occupied by the
vertex vectors~$\vec{r}_i$.  The vertices then divide according to which
side of this plane their vectors fall on.  Finding the maximum of the
modularity is then a matter of choosing this bisecting plane to maximize
the magnitude of~$\vec{R}_1$.

\begin{figure}
\begin{center}
\includegraphics[width=8cm]{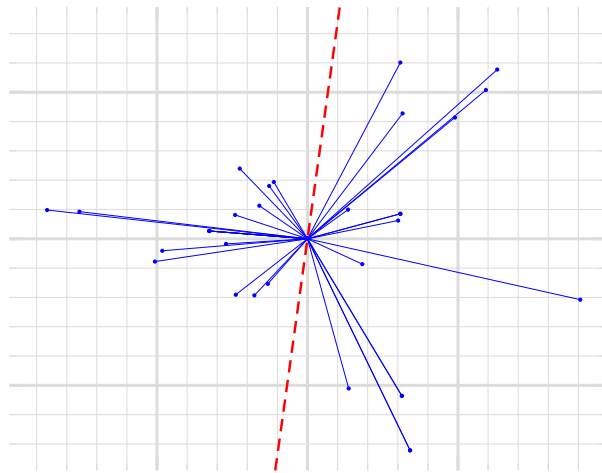}
\end{center}
\caption{A plot of the vertex vectors~$\vec{r}_i$ for a small network with
$p=2$.  The dotted line represents one of the $n$ possible topologically
distinct cut planes.}
\label{zachvecs}
\end{figure}

In general, this still leaves us with a moderately difficult optimization
problem: the number of bisecting planes that give distinct partitions of
the vertex vectors is large and difficult to enumerate as the dimension~$p$
of the space becomes large.  For the case $p=2$, however, a relatively
simple solution exists.  Consider Fig.~\ref{zachvecs}, which shows a
typical example of the vertex vectors.\footnote{In fact, this figure shows
the vectors for the ``karate club'' network used previously as an example
in Ref.~\cite{Newman06b}.}  In this two-dimensional case, there are only
$n$ topologically distinct choices of the bisecting plane (actually just a
line in this case, denoted by the dashed line in the figure), and
furthermore the divisions of the vertices that these choices represent
change by only a single vertex at a time as we rotate the plane about the
origin.  This makes it computationally simple to perform the rotation, keep
track of the value of~$\vec{R}_1$, and so find the maximum of the
modularity within this approximation.  Evaluating the magnitude of
$\vec{R}_1$ involves a constant number of operations each time we move the
line, and hence the total work involved in finding the maximum is $\Ord(n)$
for all $n$ possible positions, which is the same as the $\Ord(n)$
operations needed to separate the vertices in the $p=1$ case.

For $p>2$, we do not know of an efficient method to enumerate exhaustively
all topologically distinct bisecting planes in the vertex vector space, and
hence we have to turn to approximate methods for solving the vector
partitioning problem.  A number of reasonable heuristics have been
described in the past.  We have found acceptable though not spectacular
results, for instance, with the ``MELO'' algorithm of~\cite{AY95}, which is
essentially a greedy algorithm in which a grouping of vectors is built up
by repeatedly adding to it the vector that makes the largest contribution
to~$Q$.

\subsection{Choice of $\alpha$}
Before implementing any of these methods, a crucial question we must answer
is what value we should choose for the parameter~$\alpha$.  By tuning this
value we can improve the accuracy of our approximation to~$Q$ as follows.

By dropping the $n-p$ most negative eigenvalues, we are in effect making an
approximation to the matrix~$\mat{B}-\alpha\mat{I}$ in which it takes not
its full value $\mat{U}(\mat{D}-\alpha\mat{I})\mat{U}^T$, but an
approximate value $\mat{U}(\mat{D}'-\alpha\mat{I}')\mat{U}^T$, where
$\mat{D}'$ and $\mat{I}'$ are the matrices $\mat{D}$ and $\mat{I}$ with the
last $n-p$ diagonal elements set to zero.  We can quantify the error this
introduces by calculating the sum of the squares of the elements of the
difference between the two matrices, which is given by
\begin{eqnarray}
\chi^2 &=& \Tr[\mat{U}(\mat{D}-\alpha\mat{I})\mat{U}^T - 
               \mat{U}(\mat{D}'-\alpha\mat{I}')\mat{U}^T]^2
           \nonumber\\
       &=& \Tr[(\mat{D}-\alpha\mat{I})-(\mat{D}'-\alpha\mat{I}')]^2
        =  \!\!\sum_{i=p+1}^n (\beta_i-\alpha)^2,\qquad{}
\end{eqnarray}
where in the second line we have made use of the fact that $\mat{U}$ is
orthogonal.

Minimizing this error by setting the derivative $\d\chi^2/\d\alpha=0$, we
find
\begin{equation}
\alpha = {1\over n-p} \sum_{i=p+1}^n \beta_i.
\end{equation}
In other words, the minimal mean square error introduced by our
approximation is achieved by setting $\alpha$ equal to the mean of the
eigenvalues that have been dropped.  The only exception is when $p=n$,
where the choice of $\alpha$ makes no difference since no approximation is
being made anyway.  In our calculations we have used $\alpha=\beta_n$ in
this case, but any choice $\alpha\ge\beta_n$ would work equally well.

\section{Implementation}
\label{implementation}
Implementation of the methods described in Section~\ref{specmod} is
straightforward.  The leading-eigenvector method of Section~\ref{singlevec}
requires us to find only the single eigenvector of the modularity
matrix~$\mat{B}$ corresponding to the most positive eigenvalue.  This is
most efficiently achieved by the direct multiplication or power method.
Starting with a trial vector, we repeatedly multiply by the modularity
matrix and---unless we are unlucky enough to have chosen another
eigenvector as our trial vector---the result will converge to the
eigenvector of the matrix having the eigenvalue of largest magnitude.  In
some cases this eigenvalue will be the most positive one, in which case our
calculation ends at this point.  In other cases the eigenvalue of largest
magnitude may be negative.  If this happens then, denoting this eigenvalue
by $\beta_n$, we calculate the shifted matrix $\mat{B}-\beta_n\mat{I}$,
which has eigenvalues $\beta_i-\beta_n$ (necessarily all nonnegative) and
the same eigenvectors as the modularity matrix itself.  Then we repeat the
power-method calculation for this new matrix and this time the eigenvalue
of largest magnitude must be $\beta_1-\beta_n$ and the corresponding
eigenvector is the one we are looking for.

For the method of Section~\ref{multivec}, we require either all of the
eigenvectors of the modularity matrix or a subset corresponding to the $p$
most positive eigenvalues.  These are most conveniently calculated using
the Lanczos method or one of its variants~\cite{Meyer00}.  The fundamental
matrix operation at the heart of the Lanczos method is again multiplication
of the matrix~$\mat{B}$ into a trial vector.

Efficient implementation of any of these methods thus rests upon our
ability to rapidly multiply an arbitrary vector~$\vec{x}$ by the modularity
matrix.  This presents a problem because the modularity matrix is dense,
and hence it appears that matrix multiplications will demand $\Ord(n^2)$
time each, where $n$ is, as before, the number of vertices in the network
(which is also the size of the matrix).  By contrast, the equivalent
calculation in standard spectral partitioning is much faster because the
Laplacian matrix is sparse, having only $\Ord(n+m)$ nonzero elements, where
$m$ is the number of edges in the network.

For the standard choice, Eq.~\eqref{defscm}, of null model used to define
the modularity, however, it turns out that we can multiply by the
modularity matrix just as fast as by the Laplacian by making use of the
special structure of the matrix.  In vector notation the modularity matrix
can in this case be written
\begin{equation}
\mat{B} = \mat{A} - {\vec{k}\vec{k}^T\over2m},
\end{equation}
where $\mat{A}$ is the adjacency matrix, Eq.~\eqref{adjacency}, and
$\vec{k}$ is the $n$-element vector whose elements are the degrees~$k_i$ of
the vertices.  Then
\begin{equation}
\mat{B}\vec{x} = \mat{A}\vec{x} - {\vec{k}^T\vec{x}\over2m}\,\vec{k}.
\end{equation}
Since the adjacency matrix is sparse, having only $\Ord(m)$ elements, the
first term can be evaluated in $\Ord(m)$ time, while the second requires us
to evaluate the inner product $\vec{k}^T\vec{x}$ only once and then
multiply it into each element of $\vec{k}$ in turn, both operations taking
$\Ord(n)$ time.  Thus the entire matrix multiplication can be completed in
$\Ord(m+n)$ time, just as with the normal Laplacian matrix.  If a shift of
the eigenvalues is required to find the most positive one, as described
above, then there is an additional term $-\beta_n\mat{I}$ in the matrix,
but this also can be multiplied into an arbitrary vector in $\Ord(n)$ time,
so again the entire operation can be completed in $\Ord(m+n)$ time.

Typically $\Ord(n)$ matrix multiplications are required for either the
power method or the Lanczos method to converge to the required eigenvalues,
and hence the calculation takes $\Ord((m+n)n)$ time overall.  In the common
case in which the network is sparse and $m\propto n$, this simplifies to
$\Ord(n^2)$.

While this is, essentially, the end of the calculation for the power
method, the Lanczos method unfortunately demands more effort to find the
eigenvectors themselves.  In fact, it takes $\Ord(n^3)$ time to find all
eigenvectors of a matrix using the Lanczos method, which is quite slow.
There are on the other hand variants of the Lanczos method (as well as
other methods entirely) that can find just a few leading eigenvectors
faster than this, which makes calculations that focus on a fixed small
number of eigenvectors preferable to ones that use all eigenvectors.  In
our calculations we have primarily concentrated on algorithms that use only
one or two eigenvectors, which typically run in time~$\Ord(n^2)$ on a
sparse network.

\subsection{Refinement of the modularity}
\label{refinement}
The methods for spectral optimization of the modularity described in
Section~\ref{specmod} are only approximate.  Indeed, the problem of
modularity optimization is formally equivalent to an instance of the
NP-hard MAX-CUT problem, so it is almost certainly the case that no
polynomial-time algorithm exists that will find the modularity optimum in
all cases.  Given that the algorithms we have described run in polynomial
time, it follows that they must fail to find the optimum in some cases, and
hence that there is room for improvement of the results.

In standard graph partitioning applications it is common to use a spectral
approach based on the graph Laplacian as a first pass at the problem of
dividing a network.  The spectral method gives a broad picture of the
general shape the division should take, but there is often room for
improvement.  Typically another algorithm, such as the Kernighan--Lin
algorithm~\cite{KL70}, which swaps vertex pairs between groups in an effort
to reduce the cut size, is used to refine this first pass, and the
resulting two-stage joint strategy gives considerably better results than
either stage on its own.

We have found that a similar joint strategy gives good results in the
present case also: the divisions found with our spectral approach can be
improved in small but significant ways by adding a refinement step akin to
the Kernighan--Lin algorithm.  As described in~\cite{Newman06b}, we take an
initial division into two communities derived, for instance, from the
leading-eigenvector method of Section~\ref{singlevec} and move single
vertices between the communities so as to increase the value of the
modularity as much as possible, with the constraint that each vertex can be
moved only once.  Repeating the whole process iteratively until no further
improvement is obtained, we find a final value of the modularity which can
improve on that derived from the spectral method alone by tens of percent
in some cases, and smaller but still significant amounts in other cases.
Although the absolute gains in modularity are not always large, we find
that this refinement step is very much worth the effort it entails, raising
the typical level of performance of our methods from the merely good to the
excellent, when compared with other algorithms.  Specific examples are
given in~\cite{Newman06b}.

It is certainly possible that other refinement strategies might also give
good results.  For instance, the ``extremal optimization'' method explored
by Duch and Arenas~\cite{DA05} for optimizing modularity could be employed
as a refinement method by using the output of our spectral division as its
starting point, rather than the random configuration used as a starting
point by Duch and Arenas.

\section{Dividing networks into more than two communities}
\label{multiway}
So far we have discussed primarily methods for dividing networks into two
communities.  Many of the networks we are concerned with, however, have
more than two communities.  How can we generalize our methods to this case?
The simplest approach is repeated division into two.  That is, we use one
of the methods described above to divide our network in two and then divide
those parts in two again, and so forth.  This approach was described
briefly in Ref.~\cite{Newman06b}.

It is important to appreciate that upon further subdividing a community
within a network into two parts, the additional contribution~$\Delta Q$ to
the modularity made by this subdivision is not given correctly if we apply
the algorithms of Section~\ref{specmod} to that community alone.  That is,
we cannot simply write down the modularity matrix for the community in
question considered as a separate graph in its own right and examine the
leading eigenvector or eigenvectors.  Instead we proceed as follows.  Let
us denote the set of vertices in the community to be divided by~$G$ and let
$n_G$ be the number of vertices within this community.  Now let $\mat{S}$
be an $n_G\times c$ index matrix denoting the subdivision of the community
into $c$ subcommunities such that
\begin{equation}
S_{ij} = \begin{cases}
           1 & \quad\text{if vertex~$i$ belongs to subcommunity~$j$,}\\
           0 & \quad\text{otherwise.}
         \end{cases}
\end{equation}
Then, following Eq.~\eqref{matq}, $\Delta Q$~is the difference between the
modularities of the network before and after subdivision of the community
thus:
\begin{eqnarray}
\Delta Q &=& \sum_{i,j\in G}\:\sum_{k=1}^c B_{ij} S_{ik} S_{jk}
             - \sum_{i,j\in G} B_{ij} \nonumber\\
         &=& \sum_{k=1}^c\:\sum_{i,j\in G} 
             \biggl [ B_{ij} - \delta_{ij} \sum_{l\in G} B_{il} \biggr]
             S_{ik} S_{jk} \nonumber\\
         &=& \Tr(\mat{S}^T\mat{B}^{(G)}\mat{S}),
\label{deltaq}
\end{eqnarray}
where $\mat{B}^{(G)}$ is an $n_G\times n_G$ generalized modularity matrix
with elements indexed by the vertex labels $i,j$ of the vertices within
group~$G$ and having values
\begin{equation}
B^{(G)}_{ij} = B_{ij} - \delta_{ij} \sum_{l\in G} B_{il},
\label{genmod}
\end{equation}
with $B_{ij}$ defined by Eq.~\eqref{defsbij}.

Equation~\eqref{deltaq} has the same form as our previous expression,
Eq.~\eqref{matq}, for the modularity of the full network, and, following
the same argument as for Eqs.~\eqref{alphaq} to~\eqref{defsrk}, we can then
show that optimization of the additional modularity contribution from
subdivision of a community can also be expressed as a vector partitioning
problem, just as before.  We can approximate this vector partitioning
problem using only the leading eigenvector as in Section~\ref{singlevec} or
using more than one vector as in Section~\ref{multivec}.  The resulting
divisions can also be optimized using a ``refinement'' stage as in
Section~\ref{refinement}, to find the best possible modularity at each
step.

Using this method we can repeatedly subdivide communities to partition
networks into smaller and smaller groups of vertices and in principle this
process could continue until the network is reduced to $n$ communities
containing only a single vertex each.  Normally, however, we stop before
this point is reached because there is no point in subdividing a community
any further if no subdivision exists that will increase the modularity of
the network as a whole.  The appropriate strategy is to calculate
explicitly the modularity contribution $\Delta Q$ at each step in the
subdivision of a network, and to decline to subdivide any community for
which the value of $\Delta Q$ is not positive.  Communities with the
property of having no subdivision that gives a positive contribution to the
modularity of the network as a whole we call \defn{indivisible}; the
strategy described here is equivalent to subdividing communities repeatedly
until every remaining community is indivisible.

\begin{figure}
\begin{center}
\includegraphics[width=7cm]{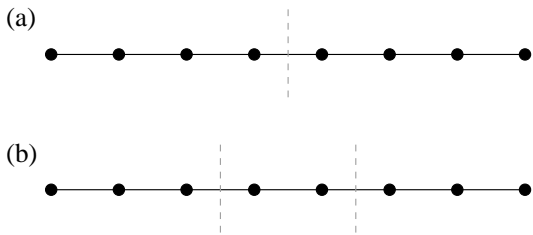}
\end{center}
\caption{Division by the method of optimal modularity of a simple network
consisting of eight vertices in a line.  (a)~The optimal division into just
two parts separates the network symmetrically into two groups of four
vertices each.  (b)~The optimal division into any number of parts divides
the network into three groups as shown here.}
\label{line8}
\end{figure}

This strategy appears to work very well in practice.  It is, however, not
perfect (a conclusion we could draw under any circumstances from the fact
that it runs in polynomial time---see above).  In particular, it is certain
that repeated subdivision of a network into two parts will in some cases
fail to find the optimal modularity configuration.  Consider, for example,
the (rather trivial) network shown in Fig.~\ref{line8}, which consists of
eight vertices connected together in a line.  By exhaustive enumeration we
can show that, among possible divisions of this network into only two
parts, the division indicated in Fig.~\ref{line8}a, right down the middle
of the line, is the one that gives the highest modularity.  The optimum
modularity over divisions into any number of parts, however, is achieved
for the three-way division shown in Fig.~\ref{line8}b.  It is clear that if
we first split the network as shown in Fig.~\ref{line8}a, no subsequent
subdivision of the network can ever find the configuration of
Fig.~\ref{line8}b, and hence our algorithm will fail in this case to find
the global optimum.  Nonetheless, the algorithm does appear to find
divisions that are close to optimal in most cases we have investigated.

Repeated subdivision is the approach we have taken to multi-community
divisions in our own work, but it is not the only possible approach.  In
some respects a more satisfying approach would be to work directly from the
expression~\eqref{finalq} for the modularity of the complete network with a
multi-community division.  Unfortunately, maximizing~\eqref{finalq}
requires us to perform a vector partitioning into more than two groups, a
problem about whose solution rather little is known.  Some general
observations are, however, worth making.  First, we note that the community
vectors~$\vec{R}_k$ in the optimal solution of a vector partitioning
problem always have directions more than $90^\circ$ apart.  To demonstrate
this, we note that the change in the contribution to Eq.~\eqref{finalq} if
we amalgamate two communities into one is
\begin{equation}
\bigl| \vec{R}_1 + \vec{R}_2 \bigr|^2 - \bigl( \bigl| \vec{R}_1 \bigr|^2
    + \bigl| \vec{R}_2 \bigr|^2 \bigr)
  = 2\vec{R}_1\cdot\vec{R}_2,
\end{equation}
which is positive if the directions of $\vec{R}_1$ and $\vec{R}_2$ are less
than $90^\circ$ apart.  Thus we can always increase the modularity by
amalgamating a pair of communities unless their vectors are more than
$90^\circ$ apart.

But the maximum number of directions more than $90^\circ$ apart that can
exist in a $p$-dimensional space is $p+1$, which means that $p+1$ is also
the maximum number of communities we can find by optimizing a
$p$-dimensional spectral approximation to the modularity.  Thus if we use
only a single eigenvector we will find at most two groups; if we use two we
will find at most three groups, and so forth.  So the choice of how many
eigenvectors~$p$ to work with is determined to some extent by the network:
if the overall optimum modularity is for a division into $c$ groups, we
will certainly fail to find that optimum if we use less than $c-1$
eigenvectors.

Second, we note that while true multi-way vector partitioning may present
problems, simple heuristics that group the vertex vectors together can
still produce good results.  For instance, White and Smyth~\cite{WS05} have
applied the standard technique of $k$-means clustering based on group
centroids to a different but related optimization problem and have found
good results.  It is possible this approach would work for our problem also
if applied to the centroids of the end-points of the vertex vectors.  It is
also possible that an intrinsically vector-based variant of $k$-means
clustering could be created to tackle the vector partitioning problem
directly, although we are not aware of such an algorithm in the current
vector partitioning literature.

\section{Negative eigenvalues and bipartite structure}
\label{negative}
It is clear from the developments of the previous sections that there is
useful information about the structure of a network stored in the
eigenvectors corresponding to the most positive eigenvalues of the
modularity matrix.  It is natural to ask whether there is also useful
information in the eigenvectors corresponding to the negative eigenvalues
and indeed it turns out that there is: the negative eigenvalues and their
eigenvectors contain information about a nontrivial type of
``anti-community structure'' that is of substantial interest in some
instances.

Consider again the case in which we divide our network into just two groups
and look once more at Eq.~\eqref{q3}, which gives the modularity in this
case.  Suppose now that instead of maximizing the terms involving the most
positive eigenvalues, we maximize the terms involving the most negative
ones.  As we can easily see from the equation, this is equivalent to
\emph{minimizing} rather than maximizing the modularity.

\begin{figure}
\begin{center}
\includegraphics[width=6cm]{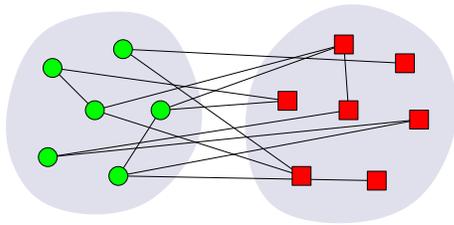}
\end{center}
\caption{A small example of an approximately bipartite network.  The
network is composed of two groups of vertices and most edges run between
vertices in different groups.}
\label{bipartex}
\end{figure}

What effect will this have on the divisions of the network that we find?
Large negative values of the modularity correspond to divisions in which
the number of edges within groups is \emph{smaller} than expected on the
basis of chance, and the number of edges between groups correspondingly
bigger.  Figure~\ref{bipartex} shows a sketch of a network having this
property.  Such networks are said to be \defn{bipartite} if there are no
edges at all within groups, or approximately bipartite if there are a few
within-group edges as in the figure.  Bipartite or approximately bipartite
graphs have attracted some attention in the recent literature.  For
instance, Kleinberg~\cite{Kleinberg99a} has suggested that small bipartite
subgraphs in the web graph may be a signature of so-called hub/authority
structure within web communities, while Holme~\etal~\cite{HLEK03} and
Estrada and Rodr{\'\i}guez-Vel\'azquez~\cite{ER05} have independently
devised measures of bipartitivity and used them to analyze a variety of
real-world networks.

The arguments above suggest that we should be able to detect bipartite or
approximately bipartite structure in networks by looking for divisions of
the vertices that minimize modularity.  In the simplest approximation, we
can do this by focusing once more on just a single term in Eq.~\eqref{q3},
that corresponding to the most negative eigenvalue~$\beta_n$, and
maximizing the coefficient of this eigenvalue by choosing $s_i=-1$ for
vertices having a negative element in the corresponding eigenvector and
$s_i=+1$ for the others.  In other words, we can achieve an approximation
to the minimum modularity division of the network by dividing vertices
according to the signs of the elements in the eigenvector~$\vec{u}_n$, and
this division should correspond roughly to the most nearly bipartite
division.  We can also append a ``refinement'' step to the calculation,
similar to that described in Section~\ref{refinement}, in which, starting
from the division given by the eigenvector, we move single vertices between
groups in an effort to minimize the modularity further.

\begin{figure}
\begin{center}
\includegraphics[width=\columnwidth]{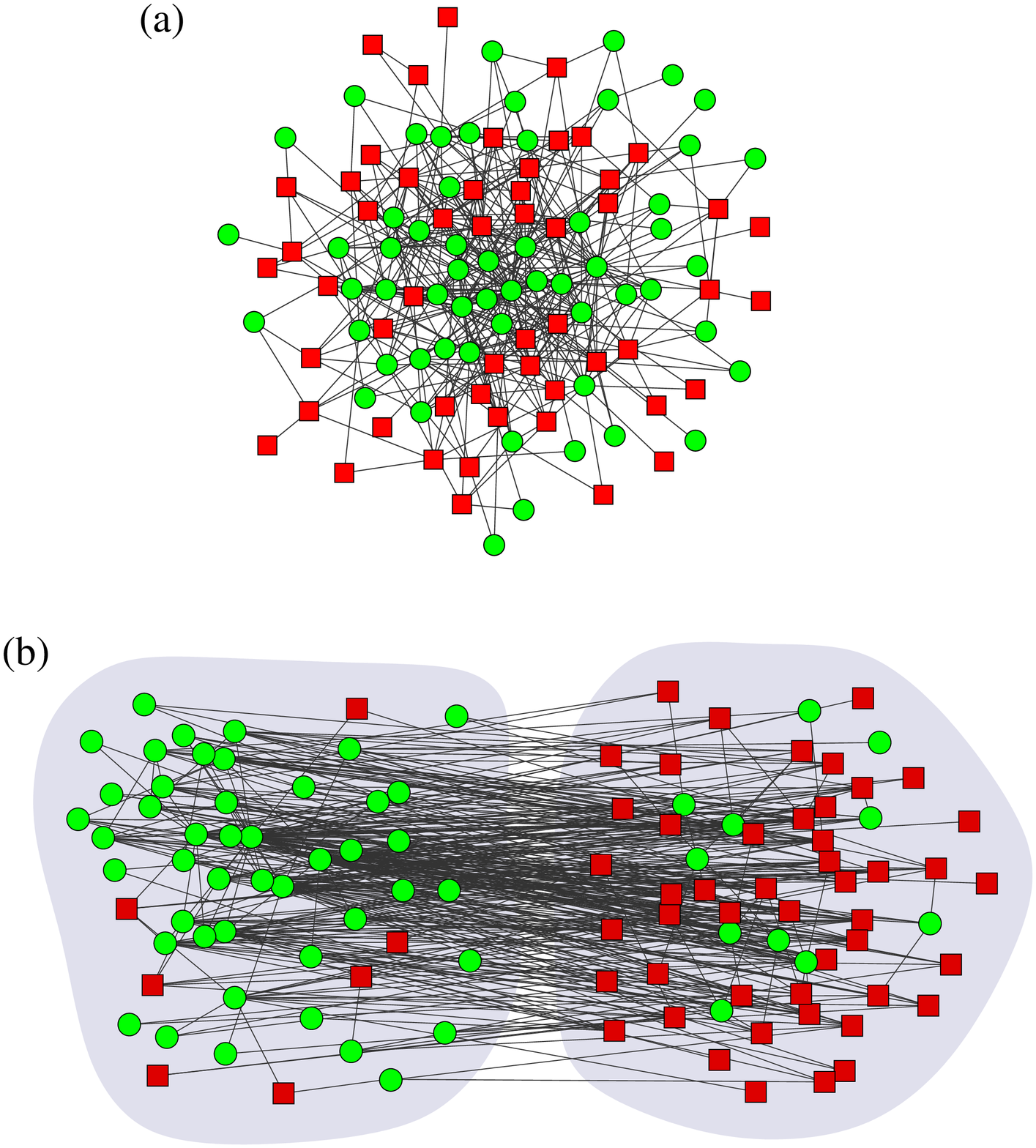}
\end{center}
\caption{(a)~The network of commonly occurring English adjectives (circles)
and nouns (squares) described in the text.  (b)~The same network redrawn
with the nodes grouped so as to minimize the modularity of the grouping.
The network is now revealed to be approximately bipartite, with one group
consisting almost entirely of adjectives and the other of nouns.}
\label{words}
\end{figure}

As an example of this type of calculation, consider Fig.~\ref{words}, which
shows a network representing juxtapositions of words in a corpus of English
text, in this case the novel \textit{David Copperfield} by Charles Dickens.
To construct this network, we have taken the 60 most commonly occurring
nouns in the novel and the 60 most commonly occurring adjectives.  (The
limit on the number of words is imposed solely to permit a clear
visualization; there is no reason in principle why the analysis could not
be extended to a much larger network.)  The vertices in the network
represent words and an edge connects any two words that appear adjacent to
one another at any point in the book.  Eight of the words never appear
adjacent to any of the others and are excluded from the network, leaving a
total of 112 vertices.

Typically adjectives occur next to nouns in English.  It is possible for
adjectives to occur next to other adjectives (``the big green bus'') or for
nouns to occur next to other nouns (``the big tour bus''), but these
juxtapositions are less common.  Thus we would expect our network to be
approximately bipartite in the sense described above: edges should run
primarily between vertices representing different types of words, with
fewer edges between vertices of the same type.  One would be hard pressed
to tell this from Fig.~\ref{words}a, however: the standard layout algorithm
used to draw the network completely fails to reveal the structure present.
Figure~\ref{words}b shows what happens when we divide the vertices by
minimizing the modularity using the method described above---a first
division according to the elements of the eigenvector with the most
negative eigenvalue, followed by a refinement stage to reduce the
modularity still further.  It is now clear that the network is in fact
nearly bipartite, and the two groups found by the algorithm correspond
closely to the known groups of adjectives and nouns, as indicated by the
shapes of the vertices.  83\% of the words are classified correctly by this
simple calculation.

Divisions with large negative modularity are---like those with large
positive modularity---not limited to having only two groups.  If we are
interested purely in minimizing the modularity we can in principle use as
many groups as we like to achieve that goal.  A division with $k$ groups is
called $k$-partite if edges run only between groups and approximately
$k$-partite if there are a few within-group edges.  One might imagine that
one could find $k$-partite structure in a network just by looking for
divisions that minimize the number of within-group edges, but brief
reflection persuades us that the optimum solution to this search problem is
always to put each vertex in a group on its own, which automatically means
that all edges lie between groups and none within groups.  As with the
ordinary community structure problem, the way to avoid this trivial
solution is to concentrate not on the total number of edges within groups
but on the difference between this number and the expected number of such
edges.  Thus, once again, we are led naturally to the consideration of
modularity as a measure of the best way to divide a network.

One way to minimize modularity over divisions into an arbitrary number of
groups is to proceed by analogy with our earlier calculations of community
structure and repeatedly divide the network in two using the
single-eigenvector method above.  Just as before, Eq.~\eqref{deltaq} gives
the additional change $\Delta Q$ in the modularity upon subdivision of a
group in a network, and the division process ends when the algorithm fails
to find any subdivision with $\Delta Q<0$.  Alternatively, one can derive
the analog of Eq.~\eqref{finalq} and thereby map the minimization of the
modularity onto a vector partitioning problem.  The appropriate definition
of the vertex vectors turns out to be
\begin{equation}
\bigl[ \vec{r}_i \bigr]_j = \sqrt{\alpha-\beta_{n+1-j}}\,U_{i,n+1-j},
\label{defsri2}
\end{equation}
where $\alpha$ is a constant chosen sufficiently large as to make
$\alpha-\beta_j\ge0$ for all terms in the sum that we keep.  Then the
modularity is given by
\begin{equation}
Q = n\alpha - \sum_{k=1}^c | \vec{R}_k |^2,
\end{equation}
with the community vectors~$\vec{R}_k$ defined according to
Eq.~\eqref{defsrk}.

\section{Other uses of the modularity matrix}
\label{otheruses}
One of the striking properties of the Laplacian matrix is that, as
described in Section~\ref{specpart}, it arises repeatedly in various
different areas of graph theory.  It is natural to ask whether the
modularity matrix also crops up in other areas.  In this section we
describe briefly two other situations in which the modularity matrix
appears, although neither has been viewed in terms of this matrix in the
past, as far as we are aware.

\subsection{Network correlations}
For our first example, suppose we have a quantity~$x_i$ defined on the
vertices~$i=1\ldots n$ of a network, such as degrees of vertices, ages of
people in a social network, numbers of hits on web pages, and so forth.
And let $\vec{x}$ be the $n$-component vector whose elements are the~$x_i$.
Then consider the quantity
\begin{equation}
r = {1\over2m} \vec{x}^T\mat{B}\vec{x},
\label{assortativity}
\end{equation}
where here we will take the same definition~\eqref{defscm} for our null
model that we have been using throughout.  Observing that $\sum_{ij}A_{ij}
= \sum_i k_i = 2m$, we can rewrite~$r$ as
\begin{eqnarray}
r &=& {1\over2m} \sum_{ij} \biggl[ A_{ij} - {k_ik_j\over2m} \biggr] x_ix_j
      \nonumber\\
  &=& {\sum_{ij} A_{ij} x_ix_j\over\sum_{ij} A_{ij}}
      - \Biggl[ {\sum_{ij} A_{ij} x_i\over\sum_{ij} A_{ij}} \Biggr]^2.
\end{eqnarray}
Note that the ratios appearing in the second line are simply averages over
all edges in the network, and hence $r$ has the form
$\av{x_ix_j}-\av{x_i}\av{x_j}$ of a correlation function measuring the
correlation of the values $x_i$ over all pairs of vertices joined by an
edge in the network.

Correlation functions of exactly this type have been considered previously
as measures of so-called ``assortative mixing,'' the tendency for adjacent
vertices in networks to have similar properties~\cite{Newman02f,Newman03c}.
For example, if the quantity $x_i$ is just the degree $k_i$ of a vertex,
then $r$~is the covariance of the degrees of adjacent vertices, which takes
positive values if vertices tend to have similar degrees to their
neighbors, high-degree vertices linking to other high-degree vertices and
low to low, and negative values if high-degree links to low.

Equation~\eqref{assortativity} is not just a curiosity, but provides some
insight concerning assortativity.  If we expand $\vec{x}$ in terms of the
eigenvectors~$\vec{u}_i$ of the modularity matrix, as we did for the
modularity itself in Eq.~\eqref{q3}, we get
\begin{equation}
r = {1\over2m} \sum_i c_i^2 \beta_i,
\end{equation}
where $\beta_i$ is again the $i$th largest eigenvalue of~$\mat{B}$ and
$c_i=\vec{u}_i^T\vec{x}$.  Thus $r$ will have a large positive value if
$\vec{x}$ has a large component in the direction of one or more of the most
positive eigenvectors of the modularity matrix, and similarly for large
negative values.  Now we recall that the leading eigenvectors of the
modularity matrix also define the communities in the network and we see
that there is a close relation between assortativity and community
structure: networks will be assortative according to some property $x$ if
the values of that property divide along the same lines as the communities
in the network.  Thus, for instance, a network will be assortative by
degree if the degrees of the vertices are partitioned such that the
high-degree vertices fall in one community and the low-degree vertices in
another.

This lends additional force to the discussion given in the introduction,
where we mentioned that different communities in networks are often found
to have different average properties such as degree.  In fact, as we now
see, this is probably the case for any property that displays significant
assortative mixing, which includes an enormous variety of quantities
measured in networks of all types.  Thus, it is not merely an observation
that different communities have different average properties---it is an
expected behavior in a network that has both community structure and
assortativity.

\subsection{Community centrality}
\label{communityc}
For our second example of other uses of the modularity matrix, we consider
centrality measures, one of the abiding interests of the network analysis
community for many decades.  In Section~\ref{singlevec} we argued that the
magnitudes of the elements of the leading eigenvector of the modularity
matrix give a measure of the ``strength'' with which vertices belong to
their assigned communities.  Thus these magnitudes define a kind of
centrality index that quantifies how central vertices are in communities.
Focusing on just a single eigenvector of the modularity matrix, however, is
limiting.  As we have seen, all the eigenvectors contain useful information
about community structure.  It is useful to ask what the appropriate
measure is of strength of community membership when the information in all
eigenvectors is taken into account.  Given Eq.~\eqref{finalq}, the obvious
candidate seems to be the projection of the vertex vector~$\vec{r}_i$ onto
the community vector~$\vec{R}_k$ of the community to which vertex~$i$
belongs.  Unfortunately, this projection depends on the arbitrary
parameter~$\alpha$, which we introduced in Eq.~\eqref{alphaq} to get around
problems caused by the negative eigenvalues of the modularity matrix.  This
in turn threatens to introduce arbitrariness into our centrality measure,
which we would prefer to avoid.  So for the purposes of defining a
centrality index we propose a slightly different formulation of the
modularity, which is less appropriate for the optimization calculations
that are the main topic of this paper, but more satisfactory for present
purposes, as we will see.

Suppose that there are $p$ positive eigenvalues of the modularity matrix
and $q$ negative ones.  We define two new sets of vertex vectors
$\set{\vec{x}_i}$ and $\set{\vec{y}_i}$, of dimension $p$ and $q$, thus:
\begin{eqnarray}
\bigl[ \vec{x}_i \bigr]_j &=& \sqrt{\beta_j}\,U_{ij},\\
\bigl[ \vec{y}_i \bigr]_j &=& \sqrt{-\beta_{n+1-j}}\,U_{i,n+1-j}.
\end{eqnarray}
(Note that $p+q<n$ since there is always at least one eigenvalue with value
zero.)  In terms of these vectors the modularity, Eq.~\eqref{matq}, can be
written as
\begin{eqnarray}
Q &=& \sum_{k=1}^c \sum_{j=1}^p \biggl[ \sum_{i=1}^n \sqrt{\beta_j}\,U_{ij}
      S_{ik} \biggr]^2 \nonumber\\
  & & {} - \sum_{k=1}^c \sum_{j=1}^q
      \biggl[ \sum_{i=1}^n \sqrt{-\beta_{n+1-j}}\,U_{i,n+1-j}
      S_{ik} \biggr]^2 \nonumber\\
  &=& \sum_{k=1}^c \sum_{j=1}^p
      \biggl[ \sum_{i\in G_k} \bigl[ \vec{x}_i \bigr]_j \biggr]^2
      - \sum_{k=1}^c \sum_{j=1}^q
      \biggl[ \sum_{i\in G_k} \bigl[ \vec{y}_i \bigr]_j \biggr]^2
      \nonumber\\
  &=& \sum_{k=1}^c | \vec{X}_k |^2 - \sum_{k=1}^c | \vec{Y}_k |^2,
\label{qxy}
\end{eqnarray}
where $G_k$ is once again the set of vertices in community~$k$ and the
community vectors $\vec{X}_k$ and $\vec{Y}_k$ are defined by
\begin{equation}
\vec{X}_k = \sum_{i\in G_k} \vec{x}_i,\qquad
\vec{Y}_k = \sum_{i\in G_k} \vec{y}_i.
\end{equation}
This reformulation avoids the use of the arbitrary constant~$\alpha$,
thereby making the vertex vectors dependent only on the network structure
and not on the way in which we choose to represent it.

Equation~\eqref{qxy} separates out the positive and negative contributions
to the modularity, the positive contributions coming from vertices that
have large corresponding elements in the eigenvectors with positive
eigenvalues, and conversely for the negative contributions.  The two
contributions correspond respectively to the traditional community
structure of Sections~\ref{modsec} and~\ref{specmod}, and to the bipartite
or $k$-partite structure discussed in Section~\ref{negative}.  It is
important to notice that while obviously the overall modularity can only be
either positive or negative, it is entirely possible for individual
vertices to simultaneously make both large positive and large negative
contributions to that modularity.  Upon reflection, this is clearly
reasonable: there is no reason why a single vertex cannot have more
connections than expected within its own community \emph{and} more
connections than expected to other communities.  In a sense,
Eq.~\eqref{qxy} may be a more fundamental representation of the modularity
than Eq.~\eqref{finalq} because it makes this separation transparent, even
if it is in practice less suitable as a basis for modularity optimization.

We can now define precisely the quantity that plays the role previously
played by the elements of the leading eigenvector in the single-eigenvector
approximation: it is the projection of $\vec{x}_i$ onto the relevant
community vector~$\vec{X}_k$, as we can see by writing the magnitude
$|\vec{X}_k|$ in Eq.~\eqref{qxy} as
\begin{equation}
|\vec{X}_k| = {\vec{X}_k^T\vec{X}_k^{\vphantom{T}}\over|\vec{X}_k|}
  = {\vec{X}_k^T\over|\vec{X}_k|} \sum_{i\in G_k} \vec{x}_i
  = \sum_{i\in G_k} \hat{\vec{X}}_k^T\vec{x}_i,
\end{equation}
where $\hat{\vec{X}}_k$ is the unit vector in the direction of~$\vec{X}_k$.
Thus each vertex vector makes a contribution to $|\vec{X}_k|$ equal to its
projection onto~$\vec{X}_k$.  In the approximation where we ignore all but
the leading eigenvector, this projection reduces to the (magnitude of) the
appropriate element of that eigenvector, as in Section~\ref{singlevec}.

The projection specifies how central vertex~$i$ is in its own community in
the traditional sense of having many connections within that community.  If
this quantity is large then we will lose a large positive contribution to
the modularity if we move the vertex to another community, which is to say
that the vertex is a strong member of its current community.

But there is also a second measure for each vertex, the projection of
$\vec{y}_i$ onto~$\vec{Y}_k$.  This projection corresponds to a more
unusual sort of centrality which is high if vertex~$i$ has many connections
to others \emph{outside} its community.  This ``outsider'' centrality
measure could also be useful in certain circumstances to identify
individuals with strong external connections.

These two projections, however, do not take precisely the form that we
expect of a centrality measure because they are functions not only of the
vertex itself (via~$\vec{x}_i$ or~$\vec{y}_i$) but also of the community in
which it is placed (via~$\vec{X}_k$ or~$\vec{Y}_k$).  Instead, therefore,
let us consider the projection in the form $|\vec{x}_i|\cos\theta_{ik}$,
where $\theta_{ik}$ is the angle between $\vec{x}_i$ and $\vec{X}_k$.  The
two parts of this expression are both of interest.  The first, the
magnitude~$|\vec{x}_i|$, measures how large a positive contribution
vertex~$i$ can \emph{potentially} make to the modularity.  The vertex only
actually makes a contribution this large if the vertex vector is aligned
with the community vector, i.e.,~if the vertex is, in a sense, ``in the
middle'' of the community to which it belongs.  Even a vertex for which
$|\vec{x}_i|$ is large may in practice make a small positive contribution
to the modularity if $\vec{x}_i$ is almost perpendicular to~$\vec{X}_k$,
i.e.,~if the vertex is ``on the edge'' of the community.

The second part of the projection, the $\cos\theta_{ik}$, is a measure
precisely of the vertex's position in the middle or on the edge of its
community.  In the parlance of social network analysis, the vertex is
either in the \defn{core} of its community ($\cos\theta_{ik}$ near~1) or in
the \defn{periphery} ($\cos\theta_{ik}$ nearer~0).  The cosine is a
property both of the vertex and of the community.

\begin{figure*}
\begin{center}
\includegraphics[clip,width=14cm]{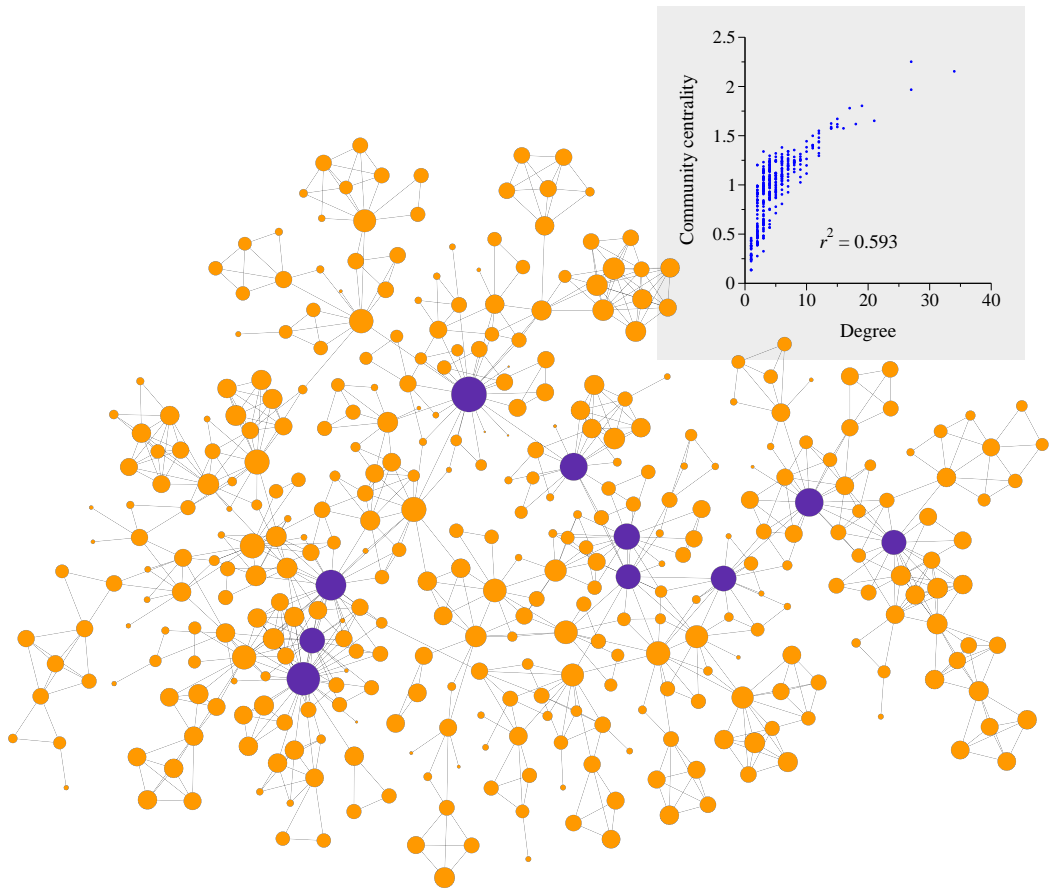}
\end{center}
\caption{A network of coauthorships between 379 scientists whose research
centers on the properties of networks of one kind or another.  Vertex
diameters indicate the community centrality and the ten vertices with
highest centralities are highlighted.  For those readers curious about the
identities of the vertices, an annotated version of this figure, names and
all, can be found at \texttt{http://www.umich.edu/\~{ }mejn/centrality}.
Inset: a scatter plot of community centrality against vertex degrees.  Like
most centrality measures, this one is correlated with degree, though only
moderately strongly.}
\label{collabs}
\end{figure*}

Let us focus here on the vector magnitudes and define two centrality
measures for vertices in a network equal to the magnitudes of the vertex
vectors $\vec{x}_i$ and~$\vec{y}_i$.  (If we prefer, we could
use~$|\vec{x}_i|^2$ instead, which is slightly easier to calculate.  If, as
is sometimes the case with centrality measures, we only care about relative
rankings of vertices, then the two are equivalent.)  These centralities are
now properties of the vertices alone and are independent of the way the
network is divided into communities.  We notice, however, that
$|\vec{x}_i|$ and $|\vec{y}_i|$ are not independent since
\begin{eqnarray}
|\vec{x}_i|^2 - |\vec{y}_i|^2
  &=& \sum_{j=1}^p \bigl( \sqrt{\beta_j}\,U_{ij} \bigr)^2 \nonumber\\
  & & {} - \sum_{j=1}^q \bigl( \sqrt{-\beta_{n+1-j}}\,U_{i,{n+1-j}} \bigr)^2
      \nonumber\\
  &=& \sum_{j=1}^n U_{ij} \beta_j U_{ji}^T
   =  B_{ii}.
\end{eqnarray}
Almost all networks considered in the literature are simple graphs,
meaning, among other things, that they have no self-edges (edges that
connect vertices to themselves) and hence $A_{ii}=0$ for all~$i$.  If the
expected number of self-edges $P_{ii}$ is also zero (as seems sensible),
then $B_{ii}=0$ and we have $|\vec{x}_i| = |\vec{y}_i|$ for all~$i$.  Thus
there is actually only one centrality for simple graphs, not two.

In fact, the choice~\eqref{defscm} for $P_{ij}$ that we and other authors
have mostly used does allow self-edges (and is in this sense slightly
unrealistic---see~\cite{MD05}), but $P_{ii}=k_i^2/2m$ is typically small
for most vertices if $m$ is large (and indeed vanishes as $m\to\infty$ if
degrees are bounded), and hence it is still true to a good approximation
that $|\vec{x}_i| \simeq |\vec{y}_i|$ and there is only one centrality.

In other words, we come to the nontrivial conclusion that the vertices with
the greatest capacity for making positive contributions to the modularity
also have the greatest capacity for making negative contributions.  The
fundamental meaning of this centrality measure is thus that there are
certain vertices that, as a consequence of their situation within the
network, have the power to make substantial contributions, either positive
or negative, to the overall modularity of the network.  For this reason, we
call this centrality measure \defn{community centrality}.  We define it to
be equal to the vector magnitude~$|\vec{x}_i|$.

An alternative way to view the community centrality is to consider how a
vertex~$i$ is situated among the other vertices in its immediate
vicinity---its neighborhood in the network.  If we were to artificially
construct a community from the vertices of this neighborhood, then that
community would presumably have a community vector~$\vec{X}_k$ with
direction close to~$\vec{x}_i$, and hence the magnitude $|\vec{x}_i|$ would
be a good measure of the actual strength with which vertex~$i$ belongs the
community.  Thus vertices with high community centrality are ones that play
a central role in their local neighborhood, regardless of where the
official community boundaries may lie.  Conversely, even when considered as
the ``center of its world'' in this way, vertex~$i$ can never play a
central role in its neighborhood in this sense if $|\vec{x}_i|$ is small.

As an example, consider Fig.~\ref{collabs}, which shows results for
community centrality for a network of coauthorships between scientists,
scientists in this case who are themselves publishing on the topic of
networks.  The network is similar to the one presented in Ref.~\cite{NG04}
but is based on more recent data, including publications up until early
2006.\footnote{The vertices of the network represent all individuals who
are authors of papers cited in the bibliographies of either of two recent
reviews on networks research~\cite{Newman03d,Boccaletti06} and edges join
every pair of individuals whose names appear together as authors of a paper
or papers in those bibliographies.  A small number of additional references
were added by hand to bring the network up to date.}  The network has a
total of 1589 scientists in it, from a broad variety of fields, but only
the 379 falling in the largest connected component are shown in the figure.
The diameters of the vertices in the figure are proportional to their
community centrality (actually to~$|\vec{x}_i|^2$---see above), and the ten
vertices having the highest centralities are highlighted.  A couple of
remarks are worth making about the results.  Without naming specific names,
we observe that all of the highlighted authors are group leaders or senior
researchers of groups working in this area.  Thus community centrality
appears to live up to its name in this admittedly anecdotal example: it
highlights those vertices that are central in their local communities.
Second, while the centrality is correlated with degree ($r^2=0.59$---see
the inset figure), the two are not perfectly correlated and in particular
some vertices have quite high centrality while having relatively low
degree.  This emphasizes the point that high centrality is an indicator of
individuals who have more connections \emph{than expected} within their
neighborhood (and hence potentially make a large contribution to the
modularity), rather than simply having a lot of connections.

\section{Conclusions}
\label{concs}
In this paper, we have studied the problem of detecting community structure
in networks.  There is already a substantial body of theory supporting the
view that community structure can be accurately quantified using the
benefit function known as modularity and hence that communities can be
detected by searching possible divisions of a network for ones that possess
high modularity.  Here we have demonstrated that the modularity can be
succinctly expressed in terms of the eigenvalues and eigenvectors of a
matrix we call the modularity matrix, which is a characteristic property of
the network and is itself independent of any division of the network into
communities.  Using this expression we have derived a series of further
results including several new and competitive algorithms for identifying
communities, a method for detecting bipartite or $k$-partite structure in
networks, and a new community centrality measure that identifies vertices
that play a central role in the communities to which they belong.

We have demonstrated a variety of applications of our methods to real-world
networks representing social, technological, and information networks.
These, however, are intended only as illustrations of the potential of
these methods.  We hope that readers will feel encouraged to apply these or
similar methods to other networks of scientific interest and we look
forward to seeing the results.

\begin{acknowledgments}
The author thanks Luis Amaral, Alex Arenas, Roger Guimer\`a, Edward
Ionides, and David Lusseau for useful and enjoyable conversations, and
Valdis Krebs and David Lusseau for providing network and other data used in
the examples.  This work was funded in part by the National Science
Foundation under grant number DMS--0405348 and by the James S. McDonnell
Foundation.
\end{acknowledgments}

\end{document}